\documentclass[11pt,A4paper]{article}
\pdfoutput=1

\usepackage{jcappub}

\usepackage{cleveref}

\usepackage{mathrsfs}
\usepackage{bbm}
\usepackage{bm}
\usepackage{dsfont}
\usepackage{tensor}
\usepackage{xspace}
\usepackage{multirow}
\usepackage{aeguill}
\usepackage{array}
\usepackage{wasysym}
\usepackage{microtype}
\usepackage{empheq}
\usepackage{ifthen}
\usepackage{amsmath}
\usepackage{subcaption}
\usepackage{wrapfig}
\usepackage{graphicx}
\graphicspath{ {./plots/} }

\newcolumntype{C}[1]{>{\centering\arraybackslash}m{#1}}

\newcommand{\ma}{\mathcal{A}}

\newcommand{\mbp}{\mathcal{B}_{\pi}}
\newcommand{\mbt}{\mathcal{B}_{t}}
\newcommand{\mbq}{\mathcal{B}_{q}}

\renewcommand\lim[2]{\underset{ #1 \rightarrow #2 }{ \mathrm{lim} } \,}

\newcommand{\delimiters}[4][]{
\ifthenelse{ \equal{#1}{1} }{  #2 #3 #4  }
					{ \ifthenelse{\equal{#1}{2}}{ \big#2 #3 \big#4 }
						{ \ifthenelse{\equal{#1}{3}}{ \Big#2 #3 \Big#4 }
							{ \ifthenelse{\equal{#1}{4}}{ \bigg#2 #3 \bigg#4 }
								{ \ifthenelse{\equal{#1}{5}}{ \Bigg#2 #3 \Bigg#4 }
									{ \left#2 #3 \right#4 }
								}
							}
						}
					}
													}

\newcommand{\Hc}{\mathcal{S}}

\newcommand\ee{\end{equation}}
\newcommand\be{\begin{equation}}
\newcommand\eea{\end{eqnarray}}
\newcommand\bea{\begin{eqnarray}}

\newlength{\boxtitlelength}
\newlength{\halfrulelength}
\newcommand{\boxtitle}[1]{\footnotesize\bf{\:#1\:}}

\definecolor{blue4}{RGB}{0,0,143}
\definecolor{red4}{RGB}{143,0,0}
\definecolor{orange}{RGB}{255,128,0}
\definecolor{darkcyan}{RGB}{0,128,128}
\definecolor{olive}{RGB}{0,128,0}
\definecolor{purple}{RGB}{128,0,128}
\definecolor{cyan2}{RGB}{0,255,255}
\definecolor{fushia}{RGB}{255,0,255}
\definecolor{mygray}{gray}{0.5}
\definecolor{lightgray}{gray}{0.85}

\def\be{\begin{equation}}
\def\ee{\end{equation}}
\def\bea{\begin{eqnarray}}
\def\eea{\end{eqnarray}}

\def\hmath$#1${\texorpdfstring{{\rmfamily\textit{#1}}}{#1}}

\makeatletter
\def\@fpheader{\relax}

\makeatother

\title{Thermodynamic coefficients in third-order relativistic fluid dynamics }

\author[a,c,d]{Teboho~A.~Moloi,}
\author[a,b,c]{Azwinndini.~Muronga}
\affiliation[a]{Department of Physics, Nelson Mandela University, Port Elizabeth, 6031, South Africa}
\affiliation[b]{Faculty of Science, Nelson Mandela University, Port Elizabeth, 6031, South Africa}
\affiliation[c]{National Institute for Theoretical and Computational Sciences (NITheCS) South Africa}
\affiliation[d]{Postgraduate School, University of Johannesburg, Cnr Kingsway $\&$ University Roads, Auckland Park Kingsway Campus, Johannesburg, 2092, South Africa}
\emailAdd{teboho.abram.moloi@gmail.com}
\emailAdd{azwinndini.muronga@gmail.com}

\abstract{We developed the third-order hydrodynamic equations using relativistic extended thermodynamics of gases with 14 independent fields. The resulting fluid equations are based on relativity principle, entropy principle and the requirement of hyperbolic, and hence finite propagation of disturbance is automatically incorporated. The expressions of entropy four-current, shear-stress tensor, dynamic pressure, and heat flux are expanded up to third-order (cubic). We explicitly present the newly calculated coefficients in the equilibrium properties of an ultra-relativistic gas regime and the non-degenerate relativistic gas. Contrary to the general cases, the non-degenerate regime eliminates fugacity from the coefficients, allowing for the easy normalization of these coefficients and the ultra-relativistic regime provides us with the upper bounds of these coefficients. We
found good agreement on some of the coefficients as compared to calculations from earlier models, specifically in kinetic theory and other coefficients had a slightly different values to those obtained in kinetic theory.}

\keywords{Dynamic pressure, relativistic thermodynamics, bulk viscosity, thermal conductivity, shear viscosity}

\date{\today}

\begin{document}

\maketitle
\flushbottom

\section{Introduction}\label{sec:intro}
Rational extended thermodynamics (RET) is an important theory that applies to non-equilibrium phenomena that are not in the local equilibrium. The symmetric-hyperbolic field equations that make up its expression are local in both space and time, and its tractable properties include the restoration of finite propagation of heat and stress disturbances. Additionally, it sheds light on the transport coefficient, shear- and bulk-viscosity, and thermal conductivity, all of which are directly linked to the gas's relaxation times. Thus, Navier–Stokes and Fourier (NSF) equations can be derived from RET.
\par
Since its inception about 200 years ago, the Navier-Stokes and Fourier (NSF) theory has been acknowledged as the preeminent theory for describing viscous flow and heat conduction in a fluid. Its practical applicability has also been repeatedly demonstrated, particularly in the engineering field. This includes geological engineering, mechanical engineering, aerospace engineering, chemical engineering, and biomedical engineering (such as thermal ablation for the treatment of thyroid nodules \cite{Chung2017}).
\par
The thermodynamics of irreversible processes (TIP) of the middle of the previous century established the non-equilibrium-thermodynamical foundation of the NSF theory \cite{Eckart,Groot1984}. The validity of TIP for situations where the presumption of local thermal equilibrium is satisfied is widely recognized. As a result, the validity range of the NSF theory is also limited. In other words, for phenomena beyond local equilibrium, the NSF theory is no longer applicable. Indeed, there exist many experimental evidences that the NSF theory becomes to be insufficient for highly non-equilibrium phenomena, for example, in shock waves and ultrasonic waves \cite{Vincent1965,Zeldovich1967}. At present, both theoretically and practically, generalization of the NSF theory is a pressing issue.
\par
In an effort to resolve TIP, a number of efforts based on the 14-field theory have been made. These include, for instance, Extended Thermodynamics in Kinetic Theory (KT) for Relativistic Gas, Degenerate and Non-Degenerate, Bosons and Permions, and for Ultra-Relativistic Case \cite{Israel_1976, Israel:1979wp,Boillat1997} and Rational Extended Thermodynamics \cite{RET,RS}, and GENERIC, which stands for General Equation for Non-Equilibrium Reversible-Irreversible Coupling \cite{Ralph2104,Matyas2021}. There has been a first attempt at comparing several non-equilibrium thermodynamics approaches in \cite{Cimmelli:2014}. In particular, RET of gases \cite{RET,RS} has been developed as a theory to investigate the very non-equilibrium phenomena indicated above, and it uses the kinetic theory as its rationale at the mesoscopic level. These theories and the numerical simulations employing multiscale approaches interact \cite{Asproulis:2008}.
\par
By deriving the quadratic fluxes of tensor (triple tensor), the entropy 4-current, shear-stress tensor, dynamic pressure, and heat flux up to cubic, we extend the work done in Liu-M{\"u}ller-Rugger (LMR) \cite{Liu:1986anpl} in this study. Now since shear-stress tensor, dynamic pressure, and heat flux are closely related to the bulk viscosity, shear viscosity, and heat conductivity, they are theoretically amenable to experiment. Considering non-degenerate and ultra-relativistic regime we solve the coefficients that appear in the entropy 4-current. We then compare this coefficients with those derived in kinetic theory. We also investigate the relation between second- and third-order coefficients. 
\par
The format of this paper is as follows: In section. \ref{sec:basiceqns}, we describe our notation, and provide the fluxes of tensor and entropy 4-current's derivation. In section. \ref{sec:dynamics}, we analyze the 14-field theory of relativistic fluid dynamics up to third-order dissipative fluxes. We compare and contrast our relativistic dissipative fluid dynamics findings with earlier findings in section. We present our results in section. \ref{sec:results}. Finally, we shall provide conclusions in section \ref{sec:con}. 


\section{Relativistic fluid dynamics preliminaries}\label{sec:basiceqns}
In this section, we review useful equations introduced by LMR \cite{Liu:1986anpl} and provide some extensions to tensor of fluxes and entropy 4-current. We start in this section,  by describing equations or formulation of relativistic fluid dynamics, specifically, the dissipative fluid in which thermodynamic tensor and heat conductivity are non-null, i.e. Expression (22.9) in ref.~\cite{Misner:1974qy}. For convenience, throughout the paper we adopt the metric $\mathrm{diag}(+,-,-,-)$, with convention which gravitational constant and speed of light are both equal to one.
\par
As a result of the primary goal of relativistic extended thermodynamics being the determination of the 14 fields. We define the fluid by 
\bea
&&V^{\alpha}(x^{\mu}), \quad \text{particle flux vector},\label{eq:fluids1}\\
&&T^{\alpha\beta}(x^{\mu}), \quad \text{energy-momentum tensor}, \quad (T^{\alpha\beta}=T^{\beta\alpha}),\label{eq:fluids2}\\
&&S^{\alpha}(x^{\mu}),\quad \text{entropy-entropy flux vector}.\label{eq:fluids3} 
\eea
Conservation laws of particle number and energy-momentum are used to determine the necessary field equations, for instance.

\bea
\partial_{\alpha}V^{\alpha} &=& 0, \label{eq:conservden} \\ 
\partial_{\alpha}T^{\alpha\beta} &=& 0, \label{eq:conservten} 
\eea
the aforementioned equations demonstrate that the rule of conservation of energy is valid. In a similar way, momentum and net charge (such as baryons) are likewise conserved.
Additionally, the flux balance is the basis for the essential field equations \cite{Liu:1986anpl, Calzetta:1997aj, Geroch:1990bw}
\bea \label{eq:diverge}
\partial_{\gamma}{A^{\alpha\beta\gamma}} = I^{\alpha\beta}, 
\eea 
where $A^{\alpha\beta\gamma}$ and $I^{\alpha\beta}$ are third order symmetric tensor and production density respectively. This two quantities are functions of variable $V^{\alpha}$ and $T^{\alpha\beta}$---they are also trace-free. Moreover, theories described by Eqns.~\eqref{eq:conservten}, \eqref{eq:conservden}, and \eqref{eq:diverge} are known as  \textbf{divergence type}, due to the fact that the left-hand-side of all this equations are 
divergences. Thus, from kinetic theory we find that quantities  $A^{\alpha\beta\gamma}$ and $I^{\alpha\beta}$ obeys the following equations
\bea \label{eq:cond}
{I^{\alpha}} _{\alpha} = 0,\quad {A^{\alpha\beta}} _{\beta} =m^{2} V^{\alpha} ,\quad A^{\alpha\beta\gamma} =A^{\alpha\gamma\beta}.
\eea 
Where $m$ is the atomic rest mass. Notice that tensor $A^{\alpha\beta\gamma}$ in this work it is totally symmetric, however, as indicated by \cite{Geroch:1990bw, Friedrichs1971aa} this is not always the case. Following their analysis which also suggest that this tensor is trace-free lead to ${A^{\alpha\beta}}_{\beta}=0 $, which is consistent with \cite{Calzetta:1997aj}. We can easily see that the new tensor in Eqn.~\eqref{eq:diverge}  simplify to the conservation law equation \eqref{eq:conservden}. Furthermore, as a result of the second law thermodynamics; the following inequality holds
\bea\label{eq:entrofluxvec}
\partial_{\alpha}S^{\alpha} \ge 0.
\eea
Thus, we close the system by introducing the constitutive equations which in our case takes the following form \cite{Liu:1986anpl,AIHPA_1997__67_2_111_0,AIHPA_1998__69_3_309_0}
\bea\label{eq:constitu}
A^{\alpha\beta\gamma} ={A} ^{\alpha\beta\gamma}(V^{\mu} , T^{\mu\nu} ), \quad I^{\alpha\beta}={I}^{\alpha\beta}(V^{\mu}, T^{\mu\nu}), 
\eea
where ${A}$ and ${I}$ are \textit{unknown} constitutive functions---if equations \eqref{eq:conservden}—\eqref{eq:diverge} and \eqref{eq:constitu} were known, they would form an explicit set of field equations. Each of these field equation solutions is referred to as a thermodynamic process. The general form of the constitutive functions $A^{\alpha\beta\gamma}(V^{\mu}, T^{\mu\nu})$, $I^{\alpha\beta}(V^{\mu}, T^{\mu\nu})$, and $S^{\alpha}(V^{\mu}, T^{\mu\nu})$ are restricted by imposing the following principles:
\begin{itemize}
  \item Principle of relativity—field equations have the same form in all frames; the constitutive functions be invariant under the change of frames.
  \item Entropy principle—the entropy density–entropy flux vector $S^{\alpha}(V^{\alpha}, T^{\alpha\beta})$ is a constitutive quantity which obeys the inequality given by Eqn.~\eqref{eq:entrofluxvec}.
  \item Requirement of hyperbolicity—ensures that Cauchy problems of our field equations are well-posed and all wave speeds are finite; our set of field equations should be symmetric hyperbolic.
\end{itemize}
In view of the fact the applications of these principles as a means of determining the structure of constitutive functions require lengthy calculations, in this paper we will refer a reader to Ref.~\cite{Liu:1986anpl} and chapter 6 in \cite{RET}.
\subsection{Constitutive relation}\label{subsec:constitutive}
Instead of using the effective notation used in Eqns.~\eqref{eq:fluids1}, \eqref{eq:fluids2} and \eqref{eq:constitu}, we follow the formalism described by \cite{Liu:1986anpl} and \cite{RET,HIST} for us to specify the linearity. Which will allow us to write net charge 4-current as \cite{Eckart}

\bea\label{eq:netcharge}
V^{\alpha} = mnu^{\alpha},
\eea
we then define the particle number density in the fluid rest frame $n \equiv \sqrt{V^{\alpha}V_{\alpha}}$. Moreover, one can show that the fluid 4-velocity is denoted by $u^{\alpha} \equiv \frac{V^{\alpha}} {{\sqrt{V^{\alpha}V_{\alpha}}}}$ such that $u^{\alpha}u_{\alpha}=1$, where $u^{\alpha}$ has three independent components. The energy-momentum tensor can be written in the following form \cite{LL}

\bea\label{eq:energym}
T^{\alpha\beta} = \rho u^{\alpha}u^{\beta}+(p(\rho,n)+\pi)h^{\alpha\beta}+2q^{(\alpha}u^{\beta)} + t^{\langle\alpha\beta\rangle},
\eea
where $\rho\equiv u_{\alpha}u_{\beta}T^{\alpha\beta}$ is the energy density in fluid rest frame, $p$ is pressure in fluid rest frame and $\pi$ is bulk viscous pressure. We then denote the projection operator of the three-space as $h^{\alpha\beta}=u^{\alpha}u^{\beta}-g^{\alpha\beta}$. Heat flux 4-current is given by $q^{\alpha}\equiv {h^{\alpha}}_{\mu}u_{\nu}T^{\mu\nu}$, it has three independent components and $t^{\langle\alpha\beta\rangle}$ is the shear stress tensor. It is worth mentioning that in Eqns.~\eqref{eq:netcharge} and \eqref{eq:energym}, $n, \rho, p$ are all functions of particle number density and the energy density which is true in a case of thermodynamics of a gas. However, this functions are naturally defined in terms of absolute temperature $T$ and fugacity $\phi$ in \cite{Liu:1986anpl,RET}) variables in statistical mechanics. The reason for considering statistical mechanics, is that the equilibrium state functions $p(\phi,T),\rho(\phi,T),n(\phi,T)$ are known. Thus, this authorize for the relativistic thermodynamics relation
\bea\label{eq:relationreltherm}
mn(\phi,T) = -\frac{1}{T}\bigg(\frac{\partial p}{\partial \phi} \bigg)_{T}, \quad \rho(\phi,T) = T^{2}\bigg(\frac{\partial }{\partial T}\bigg( \frac{p}{T}\bigg)\bigg)_{\phi}.
\eea
Furthermore, the entropy 4-current (entropy-entropy flux vector) equilibrium in statistics mechanics is well-known which reads
\bea\label{eq:fluxentr}
S^{\alpha} \equiv S|_{E} = \bigg(\frac{\partial  p}{\partial T}\bigg)_{\phi} - \frac{\phi}{T}\bigg(\frac{\partial p}{\partial \phi}\bigg)_{T}.
\eea
We make a note that fugacity and chemical potential $\mu$ differ  by a factor of $-\frac{1}{T}$, such that $\phi = -\frac{\mu}{T}$. We also mention that while the inverse functions $\phi(n,\rho)$ and $T(n,\rho)$ are exceedingly difficult to describe analytically, they are simple to compute numerically.
\par
We now consider the change of variables, i.e $V^{\alpha},\, T^{\alpha\beta}$ to $u^{\alpha},\phi,T,q^{\alpha},\pi,t^{\langle\alpha\beta\rangle}$ for convenience reasons and also because the new set if terms allow us to  identify the non-equilibrium terms with ease, while $\pi, t^{\langle\alpha\beta\rangle},q^{\alpha}$ terms vanishes in equilibrium. Considering this new set of variables, we need to provide the expression  for the flux production $I^{\alpha\beta}$, flux tensor $A^{\alpha\beta\gamma}$ and entropy 4-current $\mathcal{S}^{\alpha}$. We start by production densities tensor, which is defined as follows 
\bea\label{eq:fluxprod}
I^{\alpha\beta} = \mbp\pi(-h^{\alpha\beta}-3u^{\alpha}u^{\beta}) + 2\mbq q^{(\alpha}u^{\beta)} + \mbt t^{\langle\alpha\beta\rangle},
\eea
the coefficients $B_{\pi}, B_{q}$, and $B_{t}$ are directly related to the bulk viscosity, heat conductivity and shear viscosity, respectively. Since in principle bulk viscosity, heat conductivity, and shear viscosity can be measured one can easily identify the unknowns coefficients  in above expression. 
\par
Next, we derive the tensor of fluxes, also known as a triple tensor $A^{\alpha\beta\gamma}$; remark this triple tensor is an extension to that given by \cite{Liu:1986anpl}, in that here, we derive a triple tensor up to the second (quadratic) order. However, it is similar to the triple tensor introduced by \cite{Muronga_2010}. In order for us to obtain this, we plug equations.~\eqref{app:matrix} into Eqn.~\eqref{app:fluxten} and then use Eqn.~\eqref{app:diffeqn} (Note that the parameters ($\lambda, \tau, \sigma, \tau^{\alpha}, \sigma^{\alpha}, \sigma^{\langle\alpha\beta\rangle} $) Eqn.~\eqref{app:matrix} remain  unchanged according to Eqn.~\eqref{app:diff1}. Then, finally the extended expression for triple tensor reads as follows
\bea\label{eq:secordten}
A^{\alpha\beta\gamma} &=& \frac{1}{2}\ma^{0}_{1}g^{(\alpha \beta}u^{\gamma)}+ \frac{1}{2}\ma^{0}_{2}(g^{(\alpha\beta}u^{\gamma)}+ 2u^{\alpha}u^{\beta}u^{\gamma}) +\frac{1}{2}\ma^{1}_{1}\pi(g^{(\alpha\beta}u^{\gamma)}+2u^{\alpha}u^{\beta}u^{\gamma}) 
\nonumber \\ &+&
 \ma^{1}_{2}(-h^{(\alpha\beta}q^{\gamma)}-5 u^{(\alpha}u^{\beta}q^{\gamma)} )+
\ma^{1}_{3}t^{(\langle\alpha\beta\rangle}u^{\gamma)} + \frac{1}{2}\ma^{2}_{1}\pi^{2}(g^{(\alpha\beta}u^{\gamma)}+2u^{\alpha}u^{\beta}u^{\gamma})
\nonumber \\ &+&
 \ma^{2}_{2}\pi(-h^{(\alpha\beta}q^{\gamma)}-5q^{(\alpha}u^{\beta}u^{\gamma)}) -
\ma^{2}_{3}(h^{(\alpha\beta}u^{\gamma)}q^{\mu}q_{\mu}+3q^{(\alpha}q^{\beta}u^{\gamma)})
\nonumber \\ &-&
 \frac{1}{2}\ma^{2}_{4}(g^{(\alpha\beta}u^{\gamma)}+2u^{\alpha}u^{\beta}u^{\gamma})q^{\mu}q_{\mu}+\ma^{2}_{5}\pi t^{(\langle\beta\gamma\rangle}u^{\alpha)}+
\ma^{2}_{6}(t^{2\langle\beta\gamma\rangle}u^{\alpha}-t^{2\langle\mu\mu\rangle}u^{\alpha}u^{\beta}u^{\gamma})
 \nonumber \\ &+& 
\frac{1}{2}
\ma^{2}_{7}(g^{(\alpha\beta}u^{\gamma)}+2u^{\alpha}u^{\beta}u^{\gamma})t^{2\langle\mu\mu\rangle} + \ma^{2}_{8}(t^{\langle\alpha\beta\rangle}q^{\gamma}-2t^{\langle\nu(\beta\rangle}u^{\gamma}u^{\alpha)}q_{\nu}) 
\nonumber \\ &+&
 \ma^{2}_{9}(t^{\langle\mu(\gamma\rangle}h^{\alpha\beta)}-5t^{\langle\gamma\mu\rangle}u^{\alpha}u^{\beta})q_{\mu},
\eea
where the coefficients $\ma^{m}_{n}$ are determined  as functions of $p$, $\Gamma_{1},\Gamma_{2}$ and $\Gamma_{3}$, which depend on the following two parameters $\phi$ and $T$, with $m = \{0,1,2\}$ indicating the order of the coefficients and $n = \{1,2,3,...,9\}$ denoting the labels of the coefficients per order of the coefficients. The explicit expressions for this coefficients are determined as follows 
\bea\label{eq:coetripletensor}
\ma^{0}_{1}&=&-nm,\quad\ma^{0}_{2} = \frac{\Gamma'_{1}}{2T}, \qquad \ma^{1}_{1} = \frac{3}{T}\frac{ \begin{bmatrix}-\dot{\Gamma}_{1} & \Gamma_{1}-\Gamma'_{1} & (\frac{3}{2}\Gamma'_{2}+2\Gamma_{2})
\\
-\ddot{p} & \dot{p}-\dot{p}' & 3(-\dot{\Gamma}_{1}+\frac{1}{2}\dot{\Gamma}'_{1})
\\
\dot{p}-\dot{p}' & p' - p'' & 3(\Gamma_{1}-\frac{3}{2}\Gamma'_{1}-\frac{1}{2}\Gamma''_{1})
\end{bmatrix} }{D^{\pi}_{1}},\nonumber \\ \ma^{1}_{2} &=& \frac{1}{2T}\frac{ \begin{bmatrix}  
\Gamma_{1} & \Gamma_{2} \\
\dot{p} & - \dot{\Gamma}_{1}
\end{bmatrix}}{D_3}, \quad \ma^{1}_{3} = -\frac{3}{2T}\frac{\Gamma_{2}}{\Gamma_{1}},
\nonumber\\
\ma^{2}_{1}&=&\frac{4}{T}\frac{ 
\begin{bmatrix}
\ddot{\Gamma}_{1} & \dot{\Gamma}_{1} & \frac{3}{4}(3\Gamma'_{3} -\frac{1}{2}\Gamma''_{3})\\
\dot{\Gamma}'_{2} & (\Gamma_{2} -\frac{3}{2}\Gamma^{''}_{2}-\frac{3}{2}\Gamma'_{2} ) &  \frac{1}{2}\Gamma^{''}_{3}
\\
-\ddot{p} & \dot{p}-\dot{p}' & 3(-\dot{\Gamma}_{1}+\frac{1}{2}\dot{\Gamma}'_{1})
\\
\dot{p}-\dot{p}' & p' - p'' & 3(\Gamma_{1}-\frac{3}{2}\Gamma'_{1}-\frac{1}{2}\Gamma''_{1})
\end{bmatrix}}{D^{\pi}_{1} },
\nonumber \\ 
\ma^{2}_{2}&=& \frac{1}{2T}\frac{ \begin{bmatrix}  
\Gamma_{2} & \Gamma_{3} 
\\
\dot{p} & - \dot{\Gamma}_{1}
\end{bmatrix}}{D_3},\quad  \ma^{2}_{3}= \frac{1}{2T}\frac{ \begin{bmatrix}  
\Gamma_{2} & \Gamma'_{1} 
\\
\dot{p} & - \dot{\Gamma}_{1}
\end{bmatrix}}{D_3}, \quad \ma^{2}_{4}=\frac{1}{2T}\frac{ \begin{bmatrix}  
\Gamma_{2} & \Gamma'_{2} 
\\
\dot{p} & - \dot{\Gamma}_{1}
\end{bmatrix}}{D_3},
\\ 
\ma^{2}_{5} &=& -\frac{\Gamma_{2}}{2T\Gamma_{1}}\frac{\begin{bmatrix}
-\ddot{p} & 3(-\dot{\Gamma}_{1}+\frac{1}{2}\dot{\Gamma}'_{1} \\ 
\dot{p}-\dot{p}' & 3(-\dot{\Gamma}_{1}+\frac{1}{2}\dot{\Gamma}'_{1})
\end{bmatrix}}{D^{\pi}_{1}} +
\frac{3\Gamma'_{3}}{4T\Gamma_{1}}\frac{\begin{bmatrix}
-\ddot{p} & \dot{p}-\dot{p}' \\ 
\dot{p}-\dot{p}' & p' - p''
\end{bmatrix}}{D^{\pi}_{1}}, \quad 
\ma^{2}_{6} = -\frac{\Gamma_{3}}{2\Gamma^{2}_{1}} + \frac{\Gamma_{2}+\Gamma'_{3}}{\Gamma^{2}_{1}}, \nonumber \\
\ma^{2}_{7} &=& -\frac{\Gamma_{3}}{2\Gamma^{2}_{1}} + \frac{\Gamma'_{3}}{\Gamma^{2}_{1}},\quad \ma^{2}_{8} = \bigg(-\frac{\Gamma_{3}}{T\Gamma_{1}}+\frac{\Gamma'_{3}}{\Gamma_{1}}\bigg)\bigg(\frac{\dot{\Gamma}_{1}}{D_{3}}+\frac{\dot{p}}{D_{3}}\bigg),
\quad \ma^{2}_{9} = \bigg(-\frac{\Gamma_{3}}{T\Gamma_{1}}+\frac{\Gamma'_{3}}{\Gamma_{1}}\bigg)\frac{\dot{p}}{D_{3}}.\nonumber
\eea
Where parameters $D_3$ and $D^{\pi}_{1}$ are defined in Appendix.~\ref{app:lcf}. We make a notice that $t^{2\langle\mu\mu\rangle}$ in Eqn.~\eqref{eq:secordten} is shortened form of the following expression $t^{\langle\alpha\beta\rangle}t_{\langle\alpha\beta\rangle}$. The triple tensor \eqref{eq:secordten} computed here in {\it linear theory} is in line with the calculated triple tensor from kinetic theory \cite{Muronga_2010}. Due to the fact that we characterized linear theory by linear relation \eqref{eq:fluxprod} and triple tensor is computed up to second order, this enable us to calculate the coefficients of linear, quadratic and cubic terms of entropy-entropy flux vector. In order to compute this, 
we make use of \eqref{app:ine2} this relation may be converted into a differential form of $S^{\alpha}$ which reads
\bea\label{eq:diff_form}
dS^{\alpha}=\xi d V^{\alpha}+\xi_{\beta}d T^{\alpha\beta}+\xi_{\beta\gamma}d A^{\alpha\beta\gamma}.
\eea
Substituting in equations \eqref{eq:netcharge}, \eqref{eq:energym}, and \eqref{eq:secordten} into \eqref{eq:diff_form} then integrating throughout lead to 
the entropy 4-current, up to third order (cubic) to be written as follows 
\bea\label{eq:entropyfluxvec}
S^{\alpha}&=& \Hc^{0}_{1}u^{\alpha}+\Hc^{1}_{1}\pi u^{\alpha} + \Hc^{1}_{2}q^{\alpha}+(\Hc^{2}_{1}\pi^{2} -\Hc^{2}_{2}q^{\mu}q_{\mu}+\Hc^{2}_{3}t^{2\langle\mu\mu\rangle})u^{\alpha}+\Hc^{2}_{4}\pi q^{\alpha}+\Hc^{2}_{5}t^{2\langle\alpha\mu\rangle}q_{\mu}\nonumber \\ &+& (\Hc^{3}_{1}\pi^{3}-\Hc^{3}_{2}\pi q^{\mu}q_{\mu}+\Hc^{3}_{3}\pi t^{2\langle\mu\mu\rangle}+\Hc^{3}_{4}q_{\mu}q_{\nu}t^{\langle\mu\nu\rangle}+\Hc^{3}_{5}t^{3\langle\mu\mu\rangle})u^{\alpha}\nonumber\\&+&
(\Hc^{3}_{6}\pi^{2}-\Hc^{3}_{7}q^{\mu}q_{\mu}+\Hc^{3}_{8}t^{2\langle\mu\mu\rangle})q^{\alpha} +\Hc^{3}_{9}\pi t^{\langle\alpha\mu\rangle}q_{\mu}+\Hc^{3}_{10}t^{\langle\mu\alpha\rangle}q_{\mu},
\eea
the term $t^{3\langle\mu\mu\rangle}$ is abbreviated version of $t^{\langle\alpha\beta\rangle}t_{\langle\alpha\gamma}t^{\gamma}_{\,\beta\rangle}$. where the coefficients $\Hc^{i}_{j}$ are determined  as functions of $\phi$, $T$ and $\ma^{m}_{n}$, with $i = \{0,1,2,3\}$ indicates the order of the coefficients and $j = \{1,2,3,...,10\}$ denotes the labels of the coefficients per order of the coefficients. The explicit expressions for this coefficients are given by
\bea\label{eq:coefEntropyflux}
\Hc^{0}_{1} &=& S|_{E}, \quad \Hc^{1}_{1} = 0,\quad \Hc^{1}_{2} = \frac{1}{T},\quad \Hc^{2}_{1}=-\ma^{1}_{1}\sigma_{\pi},\quad \Hc^{2}_{2}=-\frac{1}{2}(\tau_{q}+10\ma^{1}_{2}\sigma_{q}),
\nonumber\\
\Hc^{2}_{3}&=&-\frac{1}{2\Gamma_{1}}\ma^{1}_{3},\quad
\Hc^{2}_{4}=-(\tau_{}-20\ma^{1}_{2}\sigma_{\pi})\equiv \tau_{q}+\frac{1}{3}\ma^{1}_{1}
\sigma_{q},\quad \Hc^{2}_{5} = -(\tau_{q}-2\ma^{1}_{3}\sigma_{q})\equiv-\ma^{1}_{2}\frac{2}{\Gamma_{1}}\nonumber\\
\Hc^{3}_{1} &=& -\frac{\ma^{2}_{1}}{3}\sigma_{\pi},\quad \Hc^{3}_{2} = 10\ma^{2}_{2}\sigma_{q}-12\ma^{2}_{3}\sigma_{\pi}+\frac{9}{2}\ma^{2}_{4}\sigma_{\pi},\quad \Hc^{3}_{3}= -2\frac{\ma^{2}_{5}}{\Gamma_{1}}+3\ma^{2}_{6}\sigma_{\pi}+3\ma^{2}_{7}\sigma_{\pi},\nonumber\\
\Hc^{3}_{4} &=& \ma^{2}_{8}\sigma_{q}+5\ma^{2}_{8}\sigma_{q},\quad \Hc^{3}_{5}=-\frac{\ma^{2}_{6}}{\Gamma_{1}}-\frac{\ma^{2}_{7}}{\Gamma_{1}},\quad \Hc^{3}_{6}=\frac{1}{6}\ma^{2}_{1}\sigma_{q}-10\ma^{2}_{2}\sigma_{\pi}
 \\
\Hc^{3}_{7} &=& 2\ma^{2}_{3}\sigma_{q}-\ma^{2}_{4}\sigma_{q},\quad 
\Hc^{3}_{8}=2\ma^{2}_{6}\sigma_{q}+\ma^{2}_{7}\sigma_{q},\quad 
\Hc^{3}_{9}=\frac{2\ma^{2}_{5}}{3}\sigma_{q}-2\ma^{2}_{8}\sigma_{\pi}+\ma^{2}_{9}\sigma_{\pi},\nonumber\\
\Hc^{3}_{10}&=&-\frac{1}{\Gamma_{1}}\ma^{2}_{8}-\frac{1}{\Gamma_{1}}\ma^{2}_{9}.\nonumber
\eea
We provide the parameters ($\sigma_{q},\sigma_{\pi},\tau_{q}
$) etc, in Appendix.~\ref{app:lcf}. The entropy 4-current is equivalent to the one presented in \cite{Muronga_2010,Younus:2019rrt}. However, the coefficients might be different since here we used constitutive quantities which allows us to express equations of extended thermodynamics in an approximate way. As pointed out in \cite{RET}, the result \eqref{eq:coetripletensor} and \eqref{eq:coefEntropyflux} appears to indicate that the four functions $p(\phi,T)$, $\Gamma_{1}(\phi,T)$, $\Gamma_{2}(\phi,T)$, and $\Gamma_{3}(\phi,T)$ are required to determine the flux tensor and entropy 4-current. This is not the case, since $\Gamma_{1}$, $\Gamma_{2}$ and $\Gamma_{3}$ are dependent on pressure $p(\phi,T)$. From \eqref{app:diffeqn} we obtain
\bea\label{eq:gammas}
3\Gamma_{1}-\frac{1}{2}\Gamma'_{1} =\dot{p}, \quad 4\Gamma_{2}-\frac{1}{2}\Gamma'_{2}=\dot{\Gamma}_{1},\quad \quad 5\Gamma_{3}-\frac{1}{2}\Gamma'_{3}=\dot{\Gamma}_{2},
\eea
from the above expression $\Gamma_{1}$, $\Gamma_{2}$ and $\Gamma_{3}$ can be related to $p$ by integration. Thus, we can conclude that $\Gamma_{1}$, $\Gamma_{2}$ and $\Gamma_{3}$  may be determined from $p(\phi,T)$.
\subsection{Equilibrium values}\label{subsec:equi}

In theory, equilibrium values of functions $V^{\alpha}$, $T^{\alpha\beta}$, $A^{\alpha\beta\gamma}$, and $S^{\alpha}$ can be measured, however in relativistic degenerate gas this is practically impossible. The simple option is to use statistical mechanics to compute this functions which are computed in  \cite{Muronga_2008}. The particle flux vector, energy-momentum tensor, triple tensor, entropy 4-current, respectively, take the following form 
\bea\label{eq:equilvalues}
 V^{\alpha}|_{E} &=& nm u ^{\alpha}=-\frac{1}{T}\dot{p}u^{\alpha},\quad T^{\alpha\beta}|_{E}=ph^{\alpha\beta}+\rho u^{\alpha}u^{\beta}= ph^{\alpha\beta}-(p-p')u^{\alpha}u^{\beta}\nonumber \\ 
 A^{\alpha\beta\gamma}|_{E}&=& \frac{1}{2}\ma^{0}_{1}g^{(\alpha \beta}u^{\gamma)}+ \frac{1}{2}\ma^{0}_{2}(g^{(\alpha\beta}u^{\gamma)}- 2u^{\alpha}u^{\beta}u^{\gamma})= \frac{\Gamma'_{1}}{2T}u^{\alpha}u^{\beta}u^{\gamma}- \frac{\Gamma_{1}}{2T}g^{(\alpha\beta}u^{\gamma)}\\
 S^{\alpha}|_{E}&=& Su^{\alpha}.\nonumber
\eea
Equilibrium (zeroth-order) by definition is a process in which production density vanish. Hence, the following conditions should hold;
\bea\label{eq:entprod}
I^{\alpha\beta}|_{E}=0,\quad \zeta|_{E}=0,\quad \pi|_{E}=0,\quad q^{\alpha}|_{E}=0,\quad \text{and} \quad  t^{\langle\alpha\beta\rangle}|_{E}= 0.
\eea
The comprehensive description of the transition between two sets of variables $(\phi,T)$ and $(n, \rho)$ in \eqref{eq:equilvalues} is described in \cite{Liu:1986anpl,RET}. The reason for us to introduce these values will become clear in the next section. As mentioned in ref.~\cite{Muronga_2008} in the ideal fluid limit one has five independent fields and five field equations.
   
\section{The relativistic field equations}\label{sec:dynamics}
In this section, we derive first-, second-, and third-order dissipative fluxes by an iteration scheme akin to the Maxwell iteration of the kinetic theory of gases. The iterates $I^{(n)\alpha\beta}$ result from 
\bea\label{eq:iterates}
V^{\alpha}_{\,,\alpha} = 0, \quad {T^{(n-1)\alpha\beta}}_{\,,\beta} = 0\quad {A^{(n-1)\alpha\beta\gamma}}_{\,,\gamma} = I^{(n)\alpha\beta},
\eea
to obtain the first iterations $I^{(1)\alpha\beta}$ we insertion of \eqref{eq:equilvalues} into \eqref {eq:iterates} and decomposition into time-like and space-like parts of the equations, results in the following sets of expressions
\bea\label{eq:timespaceder}
0 &=&-\ddot{p} \frac{d\alpha}{d\tau}-(\dot{p}'-\dot{p})\frac{d\ln{T}}{d\tau}-\dot{p}u^{\alpha}_{\,,\alpha}, \nonumber \\
0 &=& (\dot{p}'-\dot{p})\frac{d\alpha}{d\tau}+(p''-p')\frac{d\ln{T}}{d\tau}+p'u^{\alpha}_{\,,\alpha}, \nonumber \\
0 &=& \dot{p}{h^{\nu}}_{\mu}\alpha,_{\nu}+p'{h^{\nu}}_{\mu}(\ln{T}),_{\nu}+p'\frac{du_{\mu}}{d\tau}, \\
-3\mbp \pi^{(1)}&=& \frac{ \dot{\Gamma}'_{1}-3\dot{\Gamma}_{1}}{2T}\frac{d\alpha}{d\tau}+\frac{\Gamma''_{1}-4\Gamma'_{1}+3\Gamma_{1} }{2T}\frac{d\ln{T}}{d\tau}+\frac{\Gamma'_{1}-\Gamma_{1}}{2T}u^{\alpha}_{\,,\alpha}, \nonumber \\
-\mbq q^{(1)}_{\mu}&=&-\frac{\dot{\Gamma}_{1}}{2T}{h^{\nu}}_{\mu}\alpha,_{\nu}-\frac{\Gamma'_{1}-\Gamma_{1}}{2T}{h^{\nu}}_{\mu}(\ln{T}),_{\nu}-\frac{\Gamma'_{1}-\Gamma_{1}}{2T}\frac{du_{\mu}}{d\tau},\nonumber \\
\mbt t^{(1)}_{\langle\mu\nu\rangle} &=&-\frac{\Gamma_{1}}{T} {h^{\beta}}_{\mu}{h^{\gamma}}_{\nu}u_{\langle\mu\nu\rangle}.\nonumber
\eea
Eliminating the derivatives $\frac{d\alpha}{d\tau}, \frac{d\ln{T}}{d\tau}$ and the derivatives ${h^{\nu}}_{\mu}\alpha,_{\nu}$ from the first three equations above and inserting the results into the last three equations, allows us to determine $\pi^{(1)}, q^{(1)}_{\mu}$, and $t^{(1)}_{\mu\nu}$. Thus we obtain
\bea\label{eq:solfirstorder}
\pi^{(1)}&=& \zeta[u^{\alpha}_{,\alpha}],\nonumber\\
q^{(1)}_{\mu}&=& \kappa\bigg[h^{\alpha}_{\mu}\bigg((\ln{T} )_{,\alpha}-{\frac{du_{\mu}}{d\tau}}\bigg)\bigg],\\
t^{(1)}_{\langle\mu\nu\rangle}&=& \eta [h^{\beta}_{\,\mu}h^{\gamma}_{\,\nu}u_{\langle\beta,\gamma\rangle}].\nonumber
\eea
These equations are the relativistic analogues to the phenomenological equations of Fourier and Navier-Stokes. They were first derived by Eckart, Landau and Lifshitz \cite{Eckart,LL}. Here $[u^{\alpha}_{,\alpha}]$,$\bigg[h^{\alpha}_{\mu}\bigg((\ln{T})_{,\alpha}-\frac{du_{\mu}}{d\tau}\bigg)\bigg]$, and $[h^{\beta}_{\,\mu}h^{\gamma}_{\,\nu}u_{\langle\beta,\gamma\rangle}]$ are thermodynamic forces, $\chi$ is the bulk viscosity, $\kappa$ is the thermal conductivity and $\eta$ is the shear viscosity. This are simple algebraic expressions of thermodynamic fluxes that may lead to acausal and unstable equations of motion. The explicit expressions for $\zeta$,$\kappa$, and $\eta$ are given in \ref{eq:transc} they where also introduced by \cite{AIHPA_1997__67_2_111_0,Liu:1986anpl,RET}.
\subsection{Maxwell second order iterations}\label{subsec:seciter}

\section{Results}\label{sec:results}

The explicit expressions of the coefficients found in the flux tensor and entropy 4-current displayed in section \ref{sec:basiceqns}, are calculated and discussed in this section in a non-degenerate and ultra-relativistic regime. To understand such coefficients, one must be familiar with a particular expression for the equation of state. In practice, it is challenging to come up with an analytical expression for that equation, especially for a relativistic gas. That equation could theoretically be determined experimentally. Once the equilibrium state functions $n(\phi,T), \rho(\phi,T), p(\phi,T)$, and $\mathcal{A}^{0}_{2}(\phi,T)$ are determined, it turns out that these coefficients have a significant restriction on their reliance on fugacity $\phi$ and temperature $T$. In fact, we can compute these equilibrium state functions using statistical thermodynamics.
\par 
For a relativistic gas, J{\"u}ttner \cite{Juttner1911-ou} showed that the equilibrium distribution function for Bosons and Fermions has the form

\begin{eqnarray}\label{eq:phaseden}
f|_E=\frac{y}{\text{exp}(\frac{m}{k_{B}}\phi+\frac{u_{\alpha}}{k_{B}T}p^{\alpha}) \pm 1}, \quad \text{or}, \quad f|_E=\frac{y}{\text{exp}\bigg(\frac{m}{k_{B}}\phi+\frac{mc^{2}}{k_{B}T}\sqrt{1+\frac{p^2}{ m^2c^2} }\bigg) \pm 1},
\end{eqnarray}
this expression is valid in the rest frame of the gas.  $p^{\alpha}$ is the four-momentum of the particles such that $p^{\alpha}p_{\alpha}=m^{2}c^{2}$, then $y$ is equal to $1/\hbar^{3}$ where $\hbar$ is Planck constant, and $k_{B}$ is Boltzmann constant. J{\"u}ttner used these equilibrium distribution functions to compute the equation of state. For non-degenerate gas he found that modified Bessel functions of the second kind to take the form
\begin{eqnarray}\label{eq:mbessel}
K_{n}(\gamma)=\int^{\infty}_{0}\cosh(n\rho)\text{exp}(-\gamma\cosh\rho)d\rho,
\end{eqnarray}

we define the approximated expression of \eqref{eq:mbessel} as follows
\bea\label{eq:ambessel}
K_{n}(\gamma)= \sqrt{\frac{\pi}{2\gamma}}\frac{1}{e^{\gamma}}\bigg[ 1+\frac{4n^{2}-1}{8\gamma}+\frac{(4n^{2}-1)(4n^{2}-9)}{2!(8\gamma)^{2}}+\cdots\bigg],
\eea
we also define the following quantities which are functions of the temperature
\bea\label{eq:enthheatc}
h= mc^{2}G, \quad \hat{h}=\gamma G, \quad \Gamma = \frac{5}{3}-\frac{5}{3}\frac{1}{\gamma}+\cdots,
\eea
where $h$ is the enthalpy per particles, {\it reduced} enthalpy per particles is denoted by $\hat{h}$, and $\Gamma$ is the ratio of the heat capacities. It is important to note that \eqref{eq:mbessel} and \eqref{eq:ambessel} are both valid for large values of $\gamma$, i.e low values of temperature.
there are functions $J_{m,n}$ that appear inside the above expression \eqref{eq:mbessel} which always have an even integer as first index and therefore they can be expressed in terms of $J_{0,n}$ by making use of recurrence formulae (see \cite{https://doi.org/10.48550/arxiv.2102.02634} for more details).  In the non-degenerate regime these functions are given by
\begin{eqnarray}\label{eq:besself}
J_{0,n}(\alpha,\gamma)= \text{exp}\bigg(\frac{m}{k_{B}}\phi\bigg)\int^{\infty}_{0}\cosh^{n}(\rho)\text{exp}(\gamma\cosh\rho)d\rho.
\end{eqnarray}
Statistical thermodynamics defines the particle flux vector, energy-momentum tensor, and the tensor fluxes as moments of the distribution function $f(x^{\alpha},p^{\alpha})$
\begin{eqnarray}\label{eq:moments}
V^{\alpha}&=& mc\int p^{\alpha} f dP,\nonumber \\
T^{\alpha\beta}&=& c\int p^{\alpha} p^{\beta} f dP\\
A^{\alpha\beta\gamma}&=& \frac{c}{m}\int p^{\alpha} p^{\beta} p^{\gamma} f dP,\nonumber
\end{eqnarray}
$dP=\sqrt{-g}/p^{0}\,dp^{1}\,dp^{2}\,dp^{3}$ is the invariant element of momentum space. The equilibrium form of the equation.~\eqref{eq:moments} is obtained by substituting $f|_{E}$ given by equation.~\eqref{eq:phaseden} into the right hand side of \eqref{eq:moments} and also substituting $V^{\alpha}$, $T^{\alpha\beta}|_{E}$, and $A^{\alpha\beta\gamma}|_{E}$ which we outlined in the previous subsection into the left hand side of, we therefore arrive at
\begin{eqnarray}\label{eq:momequil}
n &=& \frac{u_{\alpha}}{c}\int p^{\alpha}f|_{E}dP,\nonumber \\
\rho &=& \frac{u_{\alpha}u_{\beta}}{c}\int p^{\alpha}p^{\beta}f|_{E}dP, \nonumber\\
p &=& \frac{c}{3}h_{\alpha\beta}\int p^{\alpha} p^{\beta} f|_{E}dP, \\
\frac{\Gamma_{1}}{T} &=& \frac{2}{3mc}h_{\alpha\beta}u_{\gamma}\int p^{\alpha}p^{\beta} p^{\gamma} f|_{E}dP.\nonumber\\
\frac{1}{2}\mathcal{A}^{0}_{1}g^{(\alpha\beta}u^{\gamma)} &+& \frac{1}{2}\mathcal{A}^{0}_{2}(g^{(\alpha\beta}u^{\gamma)} -2u^{\alpha}u^{\beta}u^{\gamma}) = \frac{c}{m}\int p^{\alpha}p^{\beta} p^{\gamma} f|_{E}dP.\nonumber
\end{eqnarray}
This set of equations allow us to compute the equilibrium state functions, with some algebra we obtain
\begin{eqnarray}\label{eq:esf}
n &=& 4\pi ym^{3}c^{3}J_{2,1}(\alpha,\gamma),\nonumber\\
\rho &=& 4\pi ym^{4}c^{5}J_{2,2}(\alpha,\gamma),\\
p &=&\frac{4}{3}\pi ym^{4}c^{5}J_{4,0}(\alpha,\gamma),\nonumber\\
\mathcal{A}^{0}_{1}&=& 4\pi m^{5}c^{3}(J_{2,1}(\alpha,\gamma)-2J_{4,1}(\alpha,\gamma).\nonumber
\end{eqnarray}
Thus, this allow us to derive the thermal equations of state which results as follows
\begin{eqnarray}\label{eq:esffinal}
\text{exp}\bigg(\frac{m}{k_{B}}\phi\bigg) &=& 4\pi y m^{2}c^{3}\frac{K_{2}}{n \gamma},\quad \gamma= \frac{mc^{2}}{k_{B}T},\nonumber\\
\rho &=& nmc^{2}\bigg(G-\frac{1}{\gamma}\bigg),\quad G\equiv\frac{K_{2}}{K_{3}},\\
p &=& nk_{B}T, \quad \text{with}, \quad n= \text{exp}\bigg(-\frac{m}{k_{B}}\phi\bigg)4\pi y m^{3}c^{3}\frac{K_{2}(\gamma)}{\gamma},\nonumber\\
\frac{\Gamma_{1}}{T} &=& nmc^{2}\frac{2}{\gamma}G,\nonumber
\end{eqnarray}
the values of the coefficients $\mathcal{A}^{m}_{n}$ for the non-degenerate relativistic gas can be derived from previous subsection.~\ref{subsec:constitutive} from flux tensor Coefficients expressions \eqref{eq:coetripletensor} by considering the non-degenerate regime after some calculations we obtain
\begin{eqnarray}\label{eq:nondegfluxtcoef}
\mathcal{A}^{0}_{1} &=& nm\bigg(1+\frac{6}{\gamma}G \bigg),
 \quad 
\mathcal{A}^{1}_{2} = -\frac{1}{\gamma}\frac{1+\frac{6}{\gamma}G-G^{2} }{1+\frac{5}{\gamma}G-G^{2} },\quad
 \mathcal{A}^{1}_{3} = \bigg(\frac{6}{\gamma}-\frac{1}{G} \bigg)\nonumber\\
\mathcal{A}^{1}_{1} &=&-\frac{6}{c^{2}}\frac{\bigg[\bigg(2-\frac{5}{\gamma^{2}}\bigg)+\bigg(\frac{19}{\gamma}-\frac{30}{\gamma^{3}}\bigg)G-\bigg(2-\frac{45}{\gamma^{2}}\bigg)G^{2}-\frac{9}{\gamma}G^{3}\bigg]}{\bigg[\frac{3}{\gamma}-\bigg(2-\frac{20}{\gamma^{2}}\bigg)G-\frac{13}{\gamma}G^{2}+2G^{3}\bigg]},
\end{eqnarray}
the remaining coefficients that appear in flux tensor, for instance, $\mathcal{A}^{2}_{1}$ through to $\mathcal{A}^{2}_{9}$ are defined in Appendix.~\ref{app:coenondegen}. Similarly, the coefficients $\mathcal{S}^{1}_{1}$ through $\mathcal{S}^{2}_{5}$ obtain from entropy-entropy flux vector coefficients \eqref{eq:coefEntropyflux} in a non-degenerate gas are given by
\begin{eqnarray}\label{eq:nondegentropycoef}
\mathcal{S}^{1}_{1} &=& 0, \quad \mathcal{S}^{1}_{2} =\frac{1}{T},\nonumber\\
\mathcal{S}^{2}_{1} &=& \frac{k_{B}}{4m^{3}c^{2}}\frac{\gamma^{2}}{n} \frac{1-\frac{1}{\gamma^{2}}+\frac{5}{\gamma}G-G^{2}}{-\bigg[\frac{3}{\gamma}-\bigg(2-\frac{20}{\gamma}\bigg)G-\frac{13}{\gamma}G^{2}+2G^{3}\bigg]},\nonumber\\
\mathcal{S}^{2}_{2} &=&\frac{k_{B}}{2m^{2}c^{6}}\frac{\gamma}{n} \frac{\frac{5}{\gamma}+\bigg(\frac{6}{\gamma^{2}} -1\bigg)G-\frac{10}{\gamma}G^{2}+G^{3}}{\bigg(1+   \frac{5}{\gamma}G -G^{2}\bigg)^{2}},\nonumber\\
\mathcal{S}^{2}_{3} &=&\frac{-k_{B}}{4m^{2}c^{4}}\frac{\gamma^{2}}{n}\frac{1}{G^{2}}\bigg(1+\frac{6}{\gamma}G\bigg),\\
\mathcal{S}^{2}_{4} &=&\frac{3k_{B}}{m^{2}c^{4}}\frac{\gamma }{n}\frac{\bigg[ -2+ \frac{5}{\gamma^{2}}-\bigg( \frac{22}{\gamma}-\frac{30}{\gamma^{3}}\bigg)G+\bigg(4- \frac{65}{\gamma^{2}}\bigg)G^{2} + \frac{22}{\gamma }G^{3} -2G^{4}\bigg]}{\bigg(1+\frac{5}{\gamma}G -G^{2}\bigg)\bigg[-\frac{3}{\gamma}+\bigg(2-\frac{20}{\gamma }\bigg)G+\frac{13}{\gamma}G^{2}-2G^{3}\bigg]},\nonumber\\
\mathcal{S}^{2}_{5} &=&\frac{k_{B}}{m^{2}c^{4}}\frac{\gamma }{n}\frac{1+\frac{6}{\gamma}G-G^{2} }{G\bigg(1+\frac{5}{\gamma}G-G^{2} \bigg)},\nonumber
\end{eqnarray}
the remaining coefficients that appear in the entropy-entropy flux vector,  for instance, $\mathcal{S}^{3}_{1}$ through to $\mathcal{S}^{3}_{10}$ are defined in Appendix.\ref{app:coenondegen}.

We numerically compute the normalized $\mathcal{S}^{i}_{j}(\gamma,G)$ coefficients and contrast them with those derived by De Groot \cite{DeGroot:1980dk} in order to demonstrate the behavior of these coefficients. We thought about things like $\mathcal{S}^{2}_{3}(\gamma,G)\times (-2nT^{2})$ or $\mathcal{S}^{2}_{1}(\gamma,G)\times 2nT^{3}$ in order for our normalization to be completely consistent with that of De Groot \cite{DeGroot:1980dk}. The benefit of first normalizing the coefficients is that it makes it possible to plot all of the coefficients without worrying about their units. Second, it makes it simple to compute both massive and massless particles. This is because if we consider massless particles and compute their coefficients numerically without normalization, the results will be undefinable. Throughout our computations, we adopt the units $\hbar=c=k_{B}=1$.
\begin{figure}[!ht]
	\centering
	\includegraphics[scale=0.8]{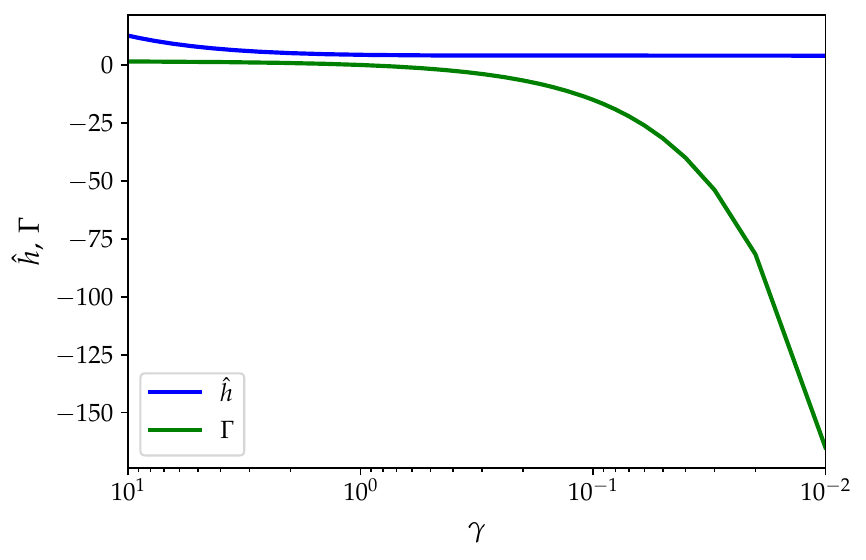}
	\caption{The reduced enthalpy per particle and ratio of heat capacities as a function of the temperature.}
	\label{fig:enthalpy}
\end{figure}	
In this context we have considered the case where a gas is characterised by large values of fugacity
\bea\label{eq:cond1}
\frac{m}{k_{B}}\phi \gg 1,
\eea
In order for us to compare our results with the known results. we focus on the upper limits of the non-degenerate gas dubbed ultra-relativistic regime, which consider cases where a gas realized a gas at very high temperatures
\bea\label{eq:cond2}
\gamma\ll 1.
\eea
The distribution function in this scenario assumes the form
\bea\label{eq:distr2}
f|_{E}=\frac{y}{\text{exp}\bigg(\frac{m}{k_B}\phi+\frac{cp}{k_{B}T}\bigg)\pm 1}
\eea
hence instead of Eqn.~\eqref{eq:esffinal} in this limit we get the following 
\bea
n &=& 4\pi y \bigg(\frac{k_{B}T}{c}\bigg)^{3} j_{2}(\phi), \quad \rho = 4\pi y c\bigg(\frac{k_{B}T}{c}\bigg)^{4} j_{3}(\phi),\\
p &=& \frac{4}{3}\pi yc \bigg(\frac{k_{B}T}{c}\bigg)^{4} j_{3}(\phi), \quad \frac{\Gamma_1}{T} = \frac{8}{3}\pi y \bigg(\frac{k_{B}T}{c}\bigg)^{5} j_{4}(\phi),
\eea
the functions $j_n(\phi)$ are defined by
\bea
j_n(\phi)=\int^{\infty}_{0} \frac{x^n}{\text{exp}\bigg(\frac{m}{k_B}\phi+x\bigg)\pm 1} dx,
\eea
and they obey the recurrence condition 
\bea
\frac{dj_n(\phi)}{d\phi}=-\frac{m}{k_B}nj_{n-1}(\phi).
\eea 

In the ultra-relativistic regime, the modified Bessel function becomes
\bea\label{eq:ultraMBf}
\stackrel{\mbox{lim}}{\gamma \to 0} K_{n}(\gamma)=\frac{2^{n-1}(n-1)!}{\gamma^{n}},
\eea
then some of $\mathcal{S}^{i}_{j}(\gamma,G)$ coefficients given by Eqn.~\eqref{eq:nondegentropycoef} after normalization and applying Eqn.~\eqref{eq:ultraMBf} reduce to just a constant, for example:
\bea\label{eq:ultralimit}
\Hc^{2}_{3} = \stackrel{\mbox{lim}}{\gamma \to 0} \frac{1}{2G^{2}}\bigg( 1+\frac{6}{\gamma}G\bigg)\equiv \stackrel{\mbox{lim}}{\gamma \to 0} \frac{1}{32}(\gamma+24)=\frac{3}{4},
\eea
similarly, 
\begin{eqnarray}\label{eq:ultrarel}
\mathcal{S}^{2}_{5}&\Rightarrow&\frac{1}{2},\quad 
\mathcal{S}^{2}_{4}\Rightarrow\frac{5}{4},\quad 
\mathcal{S}^{2}_{1}\Rightarrow\frac{3}{160},\quad
\mathcal{S}^{2}_{2}\Rightarrow \frac{9}{4},\quad \text{etc}.
\end{eqnarray}
\begin{table}[!h]
\begin{center}
    \begin{tabular}{ | C{2cm} | C{2cm} | C{2cm} | C{2cm} |}
     \hline
 \multicolumn{4}{|c|}{Ultra-relativistic limit} \\
    \hline
    Coefficients & Kinetic Theory \cite{DeGroot:1980dk} & Kinetic Theory \cite{fhumulani2010,Younus:2019rrt}& Constitutive Theory \\ \hline
    $\mathcal{S}^{2}_{1}$ &  & $\infty$ & $\frac{3}{160}$ \\ \hline
     $\mathcal{S}^{2}_{2}$ &  & $\frac{5}{4}p^{-1}$  & $\frac{9}{4}$ \\ \hline
     $\mathcal{S}^{2}_{3}$, ($\gamma'''$ )& $\frac{3}{4}$ & $\frac{3}{4}p^{-1}$ & $\frac{3}{4}$ \\ \hline
    $\mathcal{S}^{2}_{4}$ & & $\infty$ & $\frac{5}{4}$\\ \hline
    $\mathcal{S}^{2}_{5}$, ($\beta'''$)  & $\frac{1}{2}$ & $\frac{1}{2}p^{-1}$ & $\frac{1}{2}$ \\ \hline
     \end{tabular}
\end{center}
  \caption{\label{tab:table1}Compares second-order coefficients in the ultra-relativistic limit determined from Kinetic Theory and Constitutive Theory.} 
\end{table}

This limits are in contrast with those obtained by \cite{DeGroot:1980dk,Younus:2019rrt}. The summary of comparison second-order coefficients in the ultra-relativistic limit determined from
Kinetic Theory and Constitutive Theory\footnote{Indicates the work present here} is shown in Table.~\ref{tab:table1}.

\begin{figure}[!ht]
	\centering
	\includegraphics[scale=0.85]{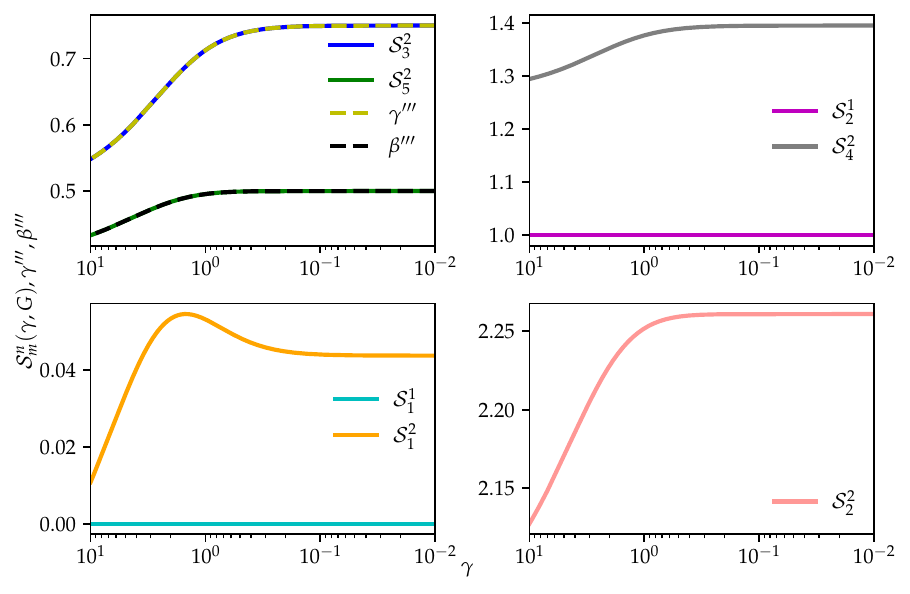}
	\caption{Plot of the first- and second-order $\Hc^{i}_{j}$ coefficients as a function of the temperature.}
	\label{fig:fscoeff}
\end{figure}	
In Fig.~\ref{fig:enthalpy} we plot the reduced enthalpy and ratio of heat capacities as a function of the temperature. Because temperature and pressure are directly proportional to enthalpy we see that if we decrease both temperature and pressure the amplitude of enthalpy also decreases. Also in Fig.~\ref{fig:enthalpy} we see that if the magnitude of ratio of heat capacities increase there is a drop in temperature.\par
\begin{figure}[!ht]
	\centering
	\includegraphics[scale=0.85]{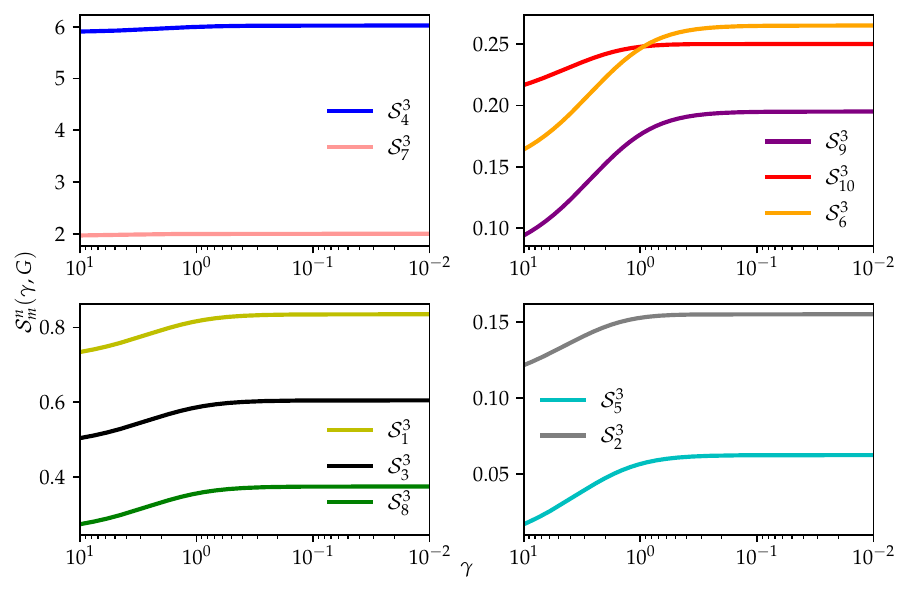}
	\caption{The temperature dependence of third-order non-classical $\Hc^{i}_{j}$ coefficients.}
	\label{fig:tcoeff}
\end{figure}	
In Fig.~\ref{fig:fscoeff} we plot all coefficients $\Hc^{i}_{j}$ up to second-order given by Eqn.~\eqref{eq:nondegentropycoef} shown with respect to the temperature. We see that both the fist-order coefficients, namely $\Hc^{1}_{1}$ and $\Hc^{1}_{2}$ remains constant through all given temperatures. In Fig.~\ref{fig:fscoeff} {\it top left panel:} we compared the coefficients $\gamma'''$, and $\beta'''$ that appears in entropy-entropy flux vector derived by De Groot \cite{DeGroot:1980dk} with the coefficients derived in this work, namely $\Hc^{2}_{3}$, and $\Hc^{2}_{5}$. We found that there is a strong agreement two the two cases analytically and computationally $\Hc^{2}_{3}$ tracks $\gamma'''$ also $\Hc^{2}_{3}$ tracks $\beta'''$ at given temperature.
\par
Also in Fig.~\ref{fig:fscoeff} {\it top right panel:} we grouped coefficients $\Hc^{1}_{2}$ and $\Hc^{2}_{4}$ since they have almost the same magnitude range, and showed the evolution of both the coefficients. In the {\it bottom left panel:} we show the behavior of coefficients $\Hc^{1}_{1}$ and $\Hc^{2}_{1}$, we notice that at $\gamma\geq1$ the magnitude of $\Hc^{2}_{1}$ increase with temperature and start to decrease with an increase of temperature at $ 1\geq\gamma\leq 0.5$, then remain constant at $\gamma\leq0.01$. We also notice that almost all second-order coefficients except $\Hc^{2}_{1}$ at $10\leq\gamma\geq1$, the magnitude of the coefficients increase with an increase of temperature (magnitude of the coefficients increase with a decrease of $\gamma$. However, at $\gamma\leq1$ the magnitude of the coefficients remains constant.
\par
Before showing the behavior third-order non-classical coefficients given by Eqn.~\eqref{app:nondegentcoe}, we also provide some of the coefficients limits in ultra-relativistic regime by making use of Eqn.~\eqref{eq:ultraMBf}, 
\bea\label{eq:ultralimit310Good}
\Hc^{3}_{10} = \stackrel{\mbox{lim}}{\gamma \to 0} \frac{18}{16\gamma}\frac{1+\frac{24}{\gamma^{2}}-\frac{20}{\gamma^{2}}-\frac{16}{\gamma^{2}} }{\bigg( \frac{4}{\gamma}\bigg)\bigg(1+\frac{24}{\gamma^{2}}-\frac{16}{\gamma^{2}}\bigg) } \equiv \frac{9}{32}\stackrel{\mbox{lim}}{\gamma \to 0} \frac{1-\frac{12}{\gamma^{2}} }{ 1+ \frac{4}{\gamma^{2}} }=\frac{9}{32},
\eea
following the same method as above we can obtain the following limits,
\bea\label{eq:ultrarel3}
\mathcal{S}^{3}_{4}&\Rightarrow & 6,\quad 
\mathcal{S}^{3}_{5}\Rightarrow\frac{3}{4},\quad 
\mathcal{S}^{3}_{7}\Rightarrow 2,\quad \text{etc}. 
\eea
\begin{table}[!h]
\begin{center}
    \begin{tabular}{ | C{2cm} | C{2cm} | C{2cm} | }
     \hline
 \multicolumn{3}{|c|}{Ultra-relativistic limit} \\
    \hline
    Coefficients & Kinetic Theory \cite{fhumulani2010,Younus:2019rrt} & Constitutive Theory \\ \hline
    $\mathcal{S}^{3}_{1}$ &  $\infty$ & $\frac{27}{32}$ \\ \hline
     $\mathcal{S}^{3}_{2}$ & $\infty$  & $\frac{1}{20}$ \\ \hline
     $\mathcal{S}^{3}_{3}$& $\infty$ & $\frac{3}{5}$ \\ \hline
    $\mathcal{S}^{3}_{4}$ & $6p^{-2}$ & $6$\\ \hline
    $\mathcal{S}^{3}_{5}$ & $\frac{3}{4}p^{-2}$ & $\frac{3}{4}$ \\ \hline
    $\mathcal{S}^{3}_{6}$ &  $\infty$ & $\frac{13}{50}$ \\ \hline
     $\mathcal{S}^{3}_{7}$ & $2p^{-2}$  & $2$ \\ \hline
     $\mathcal{S}^{3}_{8}$& $\frac{19}{50}p^{-2}$ & $\frac{3}{4}$ \\ \hline
    $\mathcal{S}^{3}_{9}$ & $\infty$ & $\frac{39}{200}$\\ \hline
    $\mathcal{S}^{3}_{10}$ & $\frac{9}{32}p^{-2}$ & $\frac{9}{32}$ \\ \hline
     \end{tabular}
\end{center}
  \caption{\label{tab:table2} Compares third-order non-classical coefficients in the ultra-relativistic limit determined from Kinetic Theory and Constitutive Theory.} 
\end{table}
This limits are in contrast with those obtained by \cite{Younus:2019rrt}. The summary of comparison second-order non-classical coefficients in the ultra-relativistic limit determined from Kinetic Theory and Constitutive Theory is shown in Table.~\ref{tab:table2}.
In Fig.~\ref{fig:tcoeff} we plot the third-order non-classical coefficients $\Hc^{i}_{j}$ given by Eqn.~\eqref{app:nondegentcoe} with respect to the temperature. For this set of coefficients in order to perform the normalization, we for example considered $\Hc^{3}_{4}\times(-2n^{2}T^{2})$. In the {\it top left panel:} we see that at $10\leq\gamma\geq1.1$, the amplitude of both the coefficients increases with an increase of temperature, but at $\gamma\leq1$ the amplitude remains constant. Both the coefficients appear to be flat, but they are not. For instance, see Fig.~\ref{fig:s37coe}. They appear to be flat because of the magnitude difference between the two coefficients. Also, in Fig.~\ref{fig:tcoeff} {\it top right panel:} we grouped coefficients $\mathcal{S}^{3}_{6},\,\mathcal{S}^{3}_{9}$, and $\mathcal{S}^{3}_{10}$ due to their magnitude and for us to be able to see the behavior of this coefficients. We see that the magnitude of coefficient $\mathcal{S}^{3}_{6}$ is smaller than that of coefficients $\mathcal{S}^{3}_{10}$ at $10\leq\gamma \geq 1.1$ and become greater at $\gamma\leq1$. In the {\it bottom left panel:} we show the evolution of coefficients $\mathcal{S}^{3}_{6},\,\mathcal{S}^{3}_{9}$, and $\mathcal{S}^{3}_{10}$ they appear to have a nice separation between them. Finally, in the {\it bottom right panel:} we plot the behavior of coefficients $\mathcal{S}^{3}_{2}$, and $\mathcal{S}^{3}_{5}$, we see both the coefficients increase with an increase of the temperature.
\par
In comparison, the second-order coefficients are normalized by multiplying the coefficients by a factor of $2pT$ or $2pT^{2}$ whereas to normalize the third-order non-classical coefficients, and we multiply by a factor of $2p^{2}$ or $2p^{2}T$. This was performed to allow our coefficients to be consistent with those obtained by \cite{DeGroot:1980dk}. We noticed that some third-order coefficients, such as $\Hc^{3}_{4}$ have a high magnitude of about six for higher temperatures, whereas the second-order coefficients don't get as high. We also establish that it seems like there might be a relation between some second-order and third-order coefficients; for example, $\Hc^{2}_{1}$ and $\Hc^{3}_{5}$ they appear to result in the same amplitude with the absolute percentage deviation of about $0.1\%$. We also notice that coefficient $\Hc^{2}_{2}$ and coefficient $\Hc^{3}_{7}$ are within $25\%$ absolute percentage difference; there might be some relation between the two coefficients. We also establish the relationship between coefficients $\Hc^{2}_{3}$ and $\Hc^{3}_{4}$ even though their amplitudes differ,  after normalization, their expression is almost similar, they differ by a square ratio of modified Bessel functions.  

\section{Conclusions}\label{sec:con}
The primary goal of this paper was to establish the third-order hydrodynamic equations employing the 14 independent fields in relativistic extended thermodynamics of gases. We were then able to derive the triple tensor up to second order and the entropy 4-current up to third-order terms in thermodynamic fluxes. We discovered that we have 10 additional non-classical coefficients in addition to those found by De-Groot and Israel-Stewart when compared to the entropy 4-current expression obtained by De-Groot and Israel-Stewart. In this paper, we called those new coefficients non-classical third order coefficients.
\par
Within the entropy 4-current the second-order and third-order terms are multiplied with coefficients whose values depend on the calculational frame, in this case we focused on non-degenerate gas values. We discovered that the coefficients become undefinable in non-degenerate gas limits when massless particles are taken into account, but they appeared to function well for massive particles. The temperature dependence of the non-classical coefficients was investigated by plotting them as functions of the temperature. To compare our third order coefficients to second order coefficients, we first reproduced the second order coefficients from De-Groot and then plotted the third order coefficients. We found very good agreement with De Groot's in kinetic theory the second-order entropy-4-current coefficients results. We also investigated the behavior of this coefficients in ultra-relativistic regime, and compared our results to those obtained by De-Groot, Fhumulani and Mohammed. We also found good agreement on some of the coefficients such as $\mathcal{S}^{2}_{3}$, and $\mathcal{S}^{2}_{5}$. Furthermore, we also evaluated the non-classical coefficients in ultra-relativistic regime, and compared our results to those obtained in kinetic theory by Fhumulani and Mohammed, we also found a good agreement on some of the non-classical coefficients such as $\mathcal{S}^{2}_{3}$, $\mathcal{S}^{3}_{4}$, $\mathcal{S}^{3}_{5}$, and $\mathcal{S}^{3}_{10}$. According to our results, third order coefficients are dependent on the preceding coefficients, i.e. second order. 
\par
By declaring the hyperbolic and thus obeying causality, we also derived the first-, second-, and third-order contributions of thermodynamic fluxes. The third-order thermodynamic fluxes are the sum of the contributions from the preceding orders. We discovered that, up to second order, these expressions agree with those obtained by  M\"{u}ller-Israel-Stewart.
\acknowledgments

This work is supported in part by Nelson Mandela University. Department of Physics under grant council, and in part by Nelson Mandela University Research Development. We would like to express our gratitude to Weigt Martin and Mohammed Younus for their insightful discussions. We would also like to thank Fhumulani Nemulodi for providing us with the soft copy of his thesis.

\appendix
\section{Constitutive function}\label{app:lcf}
Liu \cite{Liu1972} developed a bright way to execute the superiority of the entropy principle, and further advances where looked at by Ruggeri \cite{Ruggeri2005}. Salazar \cite{Salazar_2020} then provided a brief introduction to Liu’s procedure for the implementation of the entropy principle and also discussed a variance of this procedure developed by Ruggeri. Moreover, these procedures employ Lagrange multipliers, which are generated by the following considerations; let the field $(\xi,\xi_{\alpha}, \xi_{\alpha\beta})$ be Lagrange multipliers with $\xi_{\alpha\beta}=\xi_{\beta\alpha}$ symmetric and $g^{\alpha\beta}\xi_{\alpha\beta} =0$, with reference to the system \eqref{eq:conservden}, \eqref{eq:conservten} coupled to the entropy inequality \eqref{eq:diverge},

\begin{eqnarray}\label{app:ine1}
{S^{\alpha}}_{,\alpha}+\xi {V^{\alpha}}_{,\alpha}+\xi_{\beta}{T^{\alpha\beta}}_{,\alpha}+\xi_{\beta\gamma}({A^{\alpha\beta\gamma}}_{,\alpha}-I^{\beta\gamma}) \ge 0, 
\end{eqnarray}
restricted to hold for all fields ($V^{\alpha},T^{\alpha\beta}$). Liu \cite{Liu1972} addressed the existence of the Lagrange multipliers ($\xi,\xi_{\beta},\xi_{\beta\gamma}$) (they might be the functions of $V^{\alpha},T^{\alpha\beta}$) that satisfies the inequality given above. Introduction a new constitutive quantity $S^{'\alpha}$ via
\begin{eqnarray}\label{app:ine2}
S^{'\alpha}={S^{\alpha}}+\xi {V^{\alpha}}+\xi_{\beta}T^{\alpha\beta}+\xi_{\beta\gamma}{A^{\alpha\beta\gamma}},
\end{eqnarray}
the Eqn.~\ref{app:ine1} can be rewritten in the following equivalent form

\begin{eqnarray}\label{app:ine3}
{S^{'\alpha}}_{,\alpha}-V^{\alpha}\xi_{,\alpha} -T^{\alpha\beta}\xi_{\beta,\alpha}-A^{\alpha\beta\gamma}\xi_{\beta\gamma,\alpha}-\xi_{\beta\gamma}I^{\beta\gamma}\ge 0, 
\end{eqnarray}
we then can make a transformation of variables between $(V^{\alpha},T^{\alpha\beta})$ and $(\xi,\xi_{\beta},\xi_{\beta\gamma}$) since they have the number of components, the transformation takes the following form we begin from 
\begin{eqnarray}\label{app:transf}
(V^{\alpha},T^{\alpha\beta})\quad \text{to}\quad (\xi,\xi_{\beta},\xi_{\beta\gamma}),
\end{eqnarray}
insertion of $S^{'\alpha}(\xi,\xi_{\beta},\xi_{\beta\gamma})$ into Eqn.~\eqref{app:ine1} and performing the indicated differentiation of $S^{'\alpha}$ we obtain the inequality
\begin{eqnarray}\label{app:ine4}
\bigg(\frac{\partial S^{'\alpha}}{\partial \xi}-V^{\alpha}\bigg)\xi_{,\alpha} + \bigg(\frac{\partial S^{'\alpha}}{\partial \xi_{\alpha}}-T^{\alpha\beta}\bigg)\xi_{\beta,\alpha}+\bigg(\frac{\partial S^{'\alpha}}{\partial \xi_{\beta\gamma}}-A^{\alpha\beta\gamma}\bigg)\xi_{\beta\gamma,\alpha}-\xi_{\alpha\beta}I^{\alpha\beta}\ge0,
\end{eqnarray}
give main results \cite{Liu:1986anpl,RET,RS,Salazar_2020}
\bea\label{app:diff1}
V^{\alpha} =\frac{\partial S'^{\alpha}}{\partial{\xi}}, \quad T^{\alpha\beta} = \frac{\partial S'^{\alpha} }{\partial \xi_{\alpha} }, \quad A^{\langle\beta\gamma\rangle\alpha} = \frac{\partial S'^{\alpha}}{\partial \xi_{\beta\gamma}} - \frac{1}{4}g^{\alpha\beta}g_{\mu\nu}\frac{\partial S'^{\alpha}}{\partial \xi_{\mu\nu}}, 
\eea
and the residual inequality on the entropy production reads;
\bea\label{app:inequalprod}
\Xi = -\xi_{\beta\gamma}I^{\beta\gamma} \ge 0,
\eea
from Eqn.~\eqref{app:diff1} we deduce that the dependence of $(V^{\alpha}, T^{\alpha\beta}, A^{\alpha\beta\gamma})$ 
on the 
Lagrange multipliers $(\xi, \xi_{\alpha}, \xi_{\alpha\beta})$ is determined by the derivatives of the vector potential $h'^{\alpha} = h'^{\alpha}(\xi, \xi_\mu, \xi_{\mu \nu})$.\\
As indicated by \cite{Salazar_2020} the significance of this 
vector potential,
in Refs, \cite{Liu:1986anpl},\cite{RET} construct  
at first an explicit
 representation 
of $h'^{\alpha}$
as function of
$(\xi, \xi_{\mu}, \xi_{\mu\nu})$
by appealing to the principle of relativity. 
As for the case
of the tensors $I^{\alpha \beta}, A^{\alpha \beta \gamma}$ and $S^{\alpha}$ treated above,
this principle
requires that 
$ S'^\alpha $ 
to behave as
an isotropic vector function 
under arbitrary coordinate transformations \cite{wangcc1969,smithgf1965,pennisi1986}. 
\par 
Now the general form of vector $S'^{\alpha}$ up to third-order is given by the representation \cite{Liu:1986anpl,RET}
\bea\label{app:vecttrac}
S'^{\alpha} &=& \bigg(\Gamma_{0}+\frac{\partial \Gamma_{1}}{\partial G_{0}}G_{1}+\frac{1}{2}\frac{\partial^{2}\Gamma_{2}}{\partial G_{0}^{2}}G_{1}^{2}+\frac{\partial \Gamma_{2}}{\partial G_{2}}G_{2}+\frac{1}{4}\Gamma_{2}H_{2}+\frac{1}{6}\frac{\partial^{2} \Gamma_{3}}{\partial G_{0}^{2}}G_{1}^{3}+\frac{\partial^{2} \Gamma_{3}}{\partial G_{0}^{2}}G_{1}G_{2}
\nonumber\\&+& \frac{\partial \Gamma_{3}}{\partial G_{0}}G_{3}+\frac{1}{4}\frac{\partial \Gamma_{3}}{\partial G_{0}}G_{1}H_{2}+\frac{1}{6}\Gamma_{3}H_{3}\bigg)\xi^{\alpha}+\bigg( \Gamma_{2}+\frac{\partial \Gamma_{3}}{\partial G_{0}}G_{1}\bigg)\xi^{2\,\alpha\beta}\xi_{\alpha}+\Gamma_{3}\xi^{3\,\alpha\beta}\xi_{\beta}\nonumber \\ &+&
\bigg(\Gamma_{1}+\frac{\partial \Gamma_{2}}{\partial G_{0}}G_{1}+\frac{1}{2}\frac{\partial^{2}\Gamma_{3}}{\partial G_{0}^{2}}G_{1}^{2}+\frac{\partial \Gamma_{3}}{\partial G_{0}}G_{2}+\frac{1}{4}\Gamma_{3}H_{2}\bigg)\xi^{\alpha\beta}\xi_{\beta},
\eea 
where the scalars $H_{i}$ and $G_{j}$, $(i=0,2,3,4)$, $(j=0,1,2,2)$ are defined as follows
\begin{equation}
  \begin{split}
    H_0 & = \xi								\\
    H_2 & = \text{tr}(\xi^2_{\alpha \beta})	\\
    H_3 & = \text{tr}(\xi^3_{\alpha \beta})	\\
    H_4 & = \text{tr}(\xi^4_{\alpha \beta})
  \end{split}
\qquad \qquad
  \begin{split}
	G_0 & = \xi^\alpha \zeta_\alpha						\\
	G_1 & = \xi^\alpha \xi_{\alpha \beta} \xi^\beta	\\
	G_2 & = \xi^\alpha \xi_{\alpha \beta}^2 \xi^\beta	\\
	G_3 & = \xi^\alpha \xi_{\alpha \beta}^3 \xi^\beta.
  \end{split}
\end{equation}
In this paper, we derive linear constitutive equation of triple tensor $A^{\alpha\beta\gamma}$ up to second-order derivatives, which provides the extension to work done by \cite{Liu:1986anpl,RET}.

\bea\label{app:linconfluxten}
A^{\alpha\beta\gamma}&=&\frac{1}{2}\bigg(\Gamma_{1}+\frac{\partial\Gamma_{2}}{\partial G_{0}}G_{1}+\frac{1}{2}\frac{\partial^{2}\Gamma_{3}}{\partial G^{2}_{0}}G^{2}_{1}+\frac{\partial\Gamma_{3}}{\partial G_{0}}G_{2}+\frac{1}{4}\Gamma_{3}H_{2}\bigg)\bigg(g^{\alpha\beta}\xi^{\gamma}+g^{\gamma\alpha}\xi^{\beta}+g^{\beta\gamma}\xi^{\alpha}
\bigg)
\nonumber \\&+& 
\frac{1}{2}\bigg(\Gamma_{2}+\frac{\partial\Gamma_{3}}{\partial G_{0}}G_{1}\bigg)\bigg(g^{\alpha\beta}\xi^{\gamma\mu}\xi_{\mu}+g^{\gamma\alpha}\xi^{\beta\mu}\xi_{\mu}+g^{\gamma\beta}\xi^{\alpha\mu}\xi_{\mu}+\xi^{\alpha\beta}\xi^{\gamma}+\xi^{\alpha\gamma}\xi^{\beta}+\xi^{\gamma\beta}\xi^{\alpha}\bigg)
\nonumber \\&+&
\bigg[\frac{\partial \Gamma_{1}}{\partial G_{0}} + \frac{\partial^{2}\Gamma_{2}}{\partial G_{0}^{2}}G_{1}+
 \frac{\partial\Gamma_{3}}{\partial G_{0}}H_{2}+ \frac{\partial^{2}\Gamma_{3}}{\partial G_{0}^{2}}G_{2} \bigg]\xi^{\alpha}\xi^{\beta}\xi^{\gamma} + \bigg(\frac{1}{2}\frac{\partial^{2}\Gamma_{3}}{\partial G_{0}^{2}}G_{2}\bigg)\xi^{2\alpha}\xi^{2\beta}\xi^{\gamma}
\nonumber\\ &+&
\frac{1}{2}\Gamma_{3}\bigg(\xi^{2\,\alpha\beta}\xi^{\gamma}+\xi^{2\,\alpha\gamma}\xi^{\beta}+\xi^{2\,\gamma\beta}\xi^{\alpha}\bigg)
 \\&+&
\frac{\partial\Gamma_{3}}{\partial G_{0}}\bigg(
\xi^{2\alpha}\xi^{2\beta}\xi^{\gamma\mu}\xi_{\mu}+\xi^{2\alpha}\xi^{2\gamma}\xi^{\beta\mu}\xi_{\mu}+\xi^{2\gamma}\xi^{2\beta}\xi^{\alpha\mu}\xi_{\mu}\bigg)\nonumber,
\eea
although in above expressions the functions $\Gamma_0,\Gamma_1,\Gamma_2,$ and $\Gamma_3$ appears explicitly, in actuality they are considered as determined by; for example $\Gamma_{0}(G_{0},\xi)$, with $\xi$ being a function of two arbitrary functions. In \cite{Liu:1986anpl,HIST,RET} they took into account the symmetries of the field $I^{\alpha\beta},A^{\alpha\beta\gamma}$ with the prerequisite that $V^{\alpha},T^{\alpha\beta},A^{\alpha\beta\gamma}$ should be linear functions of $\xi^{\alpha\beta}$. The beauty of these treatment is that it fix functions $\Gamma_1, \Gamma_2, \Gamma_3$ in terms of $\Gamma_{0}(G_0,\xi)$.
As a result of nonlinear tensor Eqn.~\eqref{app:linconfluxten} are totally symmetric and the trace-condition Eqn.~\eqref{eq:cond} will be satisfied if the following differential equations holds.
\bea\label{app:diffeqn}
\frac{\partial }{\partial G_{0}}(G_{0}^{3}\Gamma_{1})= \frac{\partial }{\partial \xi}(G_{0}^{2}\Gamma_{0}),\quad
\frac{\partial }{\partial G_{0}}(G_{0}^{4}\Gamma_{2})= \frac{\partial }{\partial \xi}(G_{0}^{3}\Gamma_{1}),\quad
\frac{\partial }{\partial G_{0}}(G_{0}^{5}\Gamma_{3})= \frac{\partial }{\partial \xi}(G_{0}^{4}\Gamma_{2}).
\eea

Insertion of \eqref{app:vecttrac} into \eqref{app:diff1}, follows that the components of $V^{\alpha}$, and $T^{\alpha\beta}$ to linear order in $\xi_{\alpha\beta}$ have the form:
\begin{eqnarray}
V^{\alpha}&=&\bigg(\frac{\partial \Gamma_{0}}{\partial\xi}+\frac{\partial^{2}\Gamma_{1}}{\partial\xi\partial G_{0}}G_{1}\bigg)\xi^{\alpha}+ \frac{\partial \Gamma_{1}}{\partial\xi}\xi^{\alpha\beta}\xi_{\beta},\label{app:fluxv}\\
T^{\alpha\beta}&=&\bigg(\Gamma_{0}+\frac{\partial \Gamma_{1}}{\partial G_{0}}G_{1}\bigg)g^{\alpha\beta}+\Gamma_{1}\xi^{\alpha\beta}+2\bigg(\frac{\partial \Gamma_{0}}{\partial G_{0}}+\frac{\partial^{2}\Gamma_{1}}{\partial G^{2}_{0}} G_{1}\bigg)\xi^{\alpha}\xi^{\beta}\nonumber \\ &+& 2\frac{\partial \Gamma_{1}}{\partial G_{0}}(\xi^{\beta}\xi^{\alpha\gamma}\xi_{\gamma} + \xi^{\alpha}\xi^{\beta\gamma}\xi_{\gamma}).\label{app:momt}
\end{eqnarray}
The above representations of ($V^{\alpha},T^{\alpha}$) are formal since they depend upon the Lagrange multipliers whose physical significance is not yet clear. So the second step in the approach of \cite{Liu:1986anpl,RET} is to express the multipliers in terms of the physically relevant quantities such as ($\rho,p,\pi,q^{\alpha},t^{\alpha\beta},n,u^{\alpha}$) which leads to 
\begin{eqnarray}
nmu^{\alpha}&=&\bigg(\frac{\partial \Gamma_{0}}{\partial\xi}+\frac{\partial^{2}\Gamma_{1}}{\partial\xi\partial G_{0}}G_{1}\bigg)\xi^{\alpha}+ \frac{\partial \Gamma_{1}}{\partial\xi}\xi^{\alpha\beta}\xi_{\beta},\nonumber\\
\rho u^{\alpha}u^{\beta}&+&(p+\pi)h^{\alpha\beta}+u^{\alpha}q^{\beta}+u^{\beta}q^{\alpha}+t^{\langle\alpha\beta\rangle}=\bigg(\Gamma_{0}+\frac{\partial \Gamma_{1}}{\partial G_{0}}G_{1}\bigg)g^{\alpha\beta}\\ &+&\Gamma_{1}\xi^{\alpha\beta}+2\bigg(\frac{\partial \Gamma_{0}}{\partial G_{0}}+\frac{\partial^{2}\Gamma_{1}}{\partial G^{2}_{0}} G_{1}\bigg)\xi^{\alpha}\xi^{\beta} + 2 \frac{\partial \Gamma_{1}}{\partial G_{0}}(\xi^{\beta}\xi^{\alpha\gamma}\xi_{\gamma} + \xi^{\alpha}\xi^{\beta\gamma}\xi_{\gamma}),\nonumber \label{app:fluxvmomt}
\end{eqnarray}
These relations are considered as a system of 14 equations relating the Lagrange multipliers to the 14 fields.
\subsection{Constitutive relation of tensor of fluxes}
In above section, we have tensor of fluxes in terms of the Lagrange multipliers, which at this point are not meaningful. Here we relate this Lagrange multipliers to linear functions $(\pi, q^{\alpha}, t^{\langle\alpha\beta\rangle})$. In summary the equilibrium values of the Lagrange multipliers are given by (detailed calculations can be found in \cite{Liu:1986anpl,RET}): 
\bea\label{app:LGequil}
(\xi)_{E} = \alpha, \quad (\xi^{\alpha})_{E}=-\frac{u^{\alpha}}{T},\quad (\xi^{\alpha\beta})_{E}= 0,
\eea
they then perturbed the Lagrange multipliers using these equilibrium values according to the following scheme:
\bea\label{app:LGpertub}
\xi &=& -\alpha + \lambda,
\quad
\xi_{\alpha} = -\frac{u_{\alpha}}{T}+\tau_{\alpha}+\tau u_{\alpha},\\
\xi_{\alpha\beta}&=& \sigma_{\langle\alpha\beta\rangle}+\sigma h_{\alpha\beta}+ u_{\alpha}\sigma_{\beta}+u_{\beta}\sigma_{\alpha}+3\sigma u_{\alpha}u_{\beta},\nonumber
\eea
where $(\lambda, \tau_{\alpha}, \tau ,\sigma_{\langle\alpha\beta\rangle}, \sigma, \sigma_{\alpha})$ are the non equilibrium parts of the Lagrange multipliers and for simplicity reasons we omitted the subscript E describing the leading equilibrium parts. Thought out this paper we set $(\Gamma_{0})_{E}=p$ as pointed out in \cite{RET} and we adopted the following notation
\bea\label{app:deriv}
X'=\frac{\partial X}{\partial \ln{T}}, \quad \dot{X} = \frac{\partial X}{\partial \alpha}.
\eea
Insertion of Eqn.~\eqref{app:LGpertub} into Eqn.~\eqref{app:fluxvmomt} provide a system of linear equations for\newline ($\lambda,\tau,\sigma,\tau^{\alpha},\sigma^{\alpha},\sigma^{\langle\alpha\beta\rangle}$) which yield
\bea
\begin{bmatrix}
\ddot{p} &\dot{p}-\dot{p}'&3(-\dot{\Gamma}_{1} -\frac{1}{2}\dot{\Gamma}'_{1}\\
-\dot{p} & - p'&\frac{3}{2}\Gamma'_{1}+\Gamma_{1}\\
\dot{p}-\dot{p}'&p' - p''&3(\Gamma_{1}-\frac{3}{2}\Gamma'_{1}+\frac{1}{2}\Gamma''_{1}) 
\end{bmatrix}
\begin{bmatrix}
\lambda\\
-T\tau\\
\sigma
\end{bmatrix}&=&\begin{bmatrix}
0\\
\pi\\
0
\end{bmatrix}\label{app:linear1}\\
\begin{bmatrix}
\dot{p}&-\dot{\Gamma}_{1}\\
p'&(\Gamma_{1} - \Gamma'_{1} ) 
\end{bmatrix}\begin{bmatrix}
-T\tau^{\alpha}\\ 
\sigma^{\alpha}
\end{bmatrix}&=& \begin{bmatrix}
0\\
q^{\alpha}
\end{bmatrix}\label{app:linear2}
\\
\Gamma_{1}\sigma^{\langle\alpha\beta\rangle} &=& t^{\langle\alpha\beta\rangle} \label{app:linear3}
\eea 
The solution to the linear equation above are given as follows;
\bea\label{app:matrix}
\lambda &=& - \frac{\begin{bmatrix}
\dot{p}-\dot{p}' & 3(-\dot{\Gamma}_{1}+\frac{1}{2}\dot{\Gamma}'_{1})
\\
p'- p'' & 3(\Gamma_{1}-\frac{3}{2}\Gamma'_{1}-\frac{3}{2}\Gamma''_{1})
\end{bmatrix}}{D^{\pi}_{1}} \pi = {\lambda}_{\pi}\pi, \quad \sigma^{\alpha} = \frac{\dot{p}}{D_{3}}q^{\alpha} = {\sigma}_{q}q^{\alpha},
\nonumber \\ 
\tau &=&-\frac{1}{T}\frac{\begin{bmatrix}
-\ddot{p} & 3(-\dot{\Gamma}_{1}+\frac{1}{2}\dot{\Gamma}'_{1})
\\
\dot{p}-\dot{p}' & 3(\Gamma_{1}-\frac{3}{2}\Gamma'_{1}+\frac{1}{2}\Gamma''_{1})
\end{bmatrix}}{D^{\pi}_{1}} \pi = {\tau}_{\pi}\pi,\quad \sigma^{\langle\alpha\beta\rangle} = \frac{1}{\Gamma_{1}}t^{\langle\alpha\beta\rangle},
\nonumber \\
\sigma &=& -\frac{1}{T}\frac{\begin{bmatrix}
-\ddot{p} & \dot{p}-\dot{p}'
 \\
\dot{p}-\dot{p}' & p'-p''
\end{bmatrix}}{D^{\pi}_{1}} \pi = {\sigma}_{\pi}\pi,
\quad
\tau^{\alpha} = -\frac{1}{T}\frac{\dot{\Gamma}_{1}}{D_{3}}q^{\alpha} = \tau_{q}q^{\alpha},
\eea
the parameters $D^{\pi}_{1}$ and $D_{3}$ are abbreviations for the determinants (the matrix involving derivatives of the equilibrium equation of state $p$ and the derivatives of $\Gamma_{1}$) of the linear system \eqref{app:linear1} and \eqref{app:linear2}, respectively , see \cite{Liu:1986anpl,RET} for more detailed information. Substituting the linearized representations of the Lagrange multipliers \eqref{app:LGpertub} which also can be found in \cite{Liu:1986anpl} into \eqref{app:linconfluxten} result in the following linear flux tensor expression
\bea\label{app:fluxten}
A^{\alpha\beta\gamma} &=& -\frac{1}{T}\bigg\{ \bigg[ -\frac{1}{2}\Gamma'_{1} + \frac{1}{2}\dot{\Gamma}'_{1}\lambda+\bigg(\frac{1}{2}\Gamma''_{1}+\frac{1}{2}\Gamma'_{1}\bigg)(-T\tau)+ 3\bigg(\frac{1}{4}\Gamma''_{2}-\Gamma'_{2}+2\Gamma_{2}\bigg)\sigma \bigg]u^{\alpha}u^{\beta}u^{\gamma}
\nonumber \\ &+& 
\bigg[ \frac{1}{2}\Gamma_{1}-\frac{1}{2}\dot{\Gamma}_{1}\lambda + \frac{1}{2}\bigg(\Gamma_{1}-\Gamma'_{1}\bigg)(-T\tau) +
\bigg(-\frac{3}{4}\Gamma'_{2}+\Gamma_{2}\bigg)\sigma \bigg](g^{\alpha\beta}u^{\gamma}+g^{\beta\gamma}u^{\alpha}+g^{\gamma\alpha}u^{\beta})
\nonumber \\ &+&
\frac{1}{2}\Gamma'_{1}T(u^{\alpha}u^{\beta}\tau^{\gamma} + u^{\beta}u^{\gamma}\tau^{\alpha} + u^{\gamma}u^{\alpha}\tau^{\beta})+ \bigg(\Gamma_{2}-\frac{1}{2}\Gamma'_{2} \bigg)(u^{\alpha}u^{\beta}\sigma^{\gamma} + u^{\beta}u^{\gamma}\sigma^{\alpha} + u^{\gamma}u^{\alpha}\sigma^{\beta})
\nonumber \\ &-&
\frac{1}{2}\Gamma_{1}T(g^{\alpha\beta}\tau^{\gamma}+g^{\beta\gamma}\tau^{\alpha}+g^{\gamma\alpha}\tau^{\beta})
+
\frac{1}{2}\Gamma_{2}(g^{\alpha\beta}\sigma^{\gamma}+g^{\beta\gamma}\sigma^{\alpha}+g^{\gamma\alpha}\sigma^{\beta})
\nonumber \\ &+&
\frac{1}{2}\Gamma_{2}(\sigma^{\langle\alpha\beta\rangle}u^{\gamma}+\sigma^{\langle\beta\gamma\rangle}u^{\alpha}+\sigma^{\langle\gamma\alpha\rangle}u^{\beta})
\nonumber \\&+&
\bigg[-\frac{1}{2}\ddot{\Gamma}'_{1}\lambda^{2} +\Gamma''_{2}\lambda(-T\tau) -2\Gamma'_{1}(T\tau)^{2} 
\nonumber \\ &+& \bigg( \frac{1}{2}\Gamma''_{2}-9\Gamma'_{2}+\Gamma_{2}\bigg)(-T\tau\sigma)+\bigg(\frac{27}{4}\Gamma'_{3}-\frac{3}{2}\Gamma''_{3} \bigg)\sigma^{2}+\frac{3}{2}\Gamma''_{\lambda\sigma}\bigg]u^{\alpha}u^{\beta}u^{\gamma}
\nonumber \\ &+&
\bigg[ \frac{1}{2}\ddot{\Gamma}_{1}\lambda^{2}+\frac{1}{2}\bigg(\dot{\Gamma}_{1}+\Gamma'_{1}\bigg)\lambda(-T\tau)+\Gamma_{2}(T\tau)^{2}+\bigg(\Gamma_{2}-\frac{3}{2}\Gamma'_{2}-\frac{3}{2}\Gamma''_{2}\bigg)(-T\tau\sigma)
\nonumber \\ &+&
\bigg(\frac{9}{4}\Gamma'_{3}-\frac{3}{8}\Gamma''_{3} \bigg)\sigma^{2}+\frac{3}{4}\Gamma''_{3}\lambda\sigma\bigg](g^{\alpha\beta}u^{\gamma}+g^{\beta\gamma}u^{\alpha}+g^{\gamma\alpha}u^{\beta})
\nonumber \\ &-&
\frac{1}{2}\Gamma_{2}(h^{\alpha\beta}u^{\gamma}+h^{\beta\gamma}u^{\alpha}+h^{\gamma\alpha}u^{\beta})\sigma^{\mu}\tau_{\mu}+\frac{1}{2}\Gamma'_{1}T^{2}(\tau^{\alpha}\tau^{\beta}u^{\gamma}+\tau^{\gamma}\tau^{\beta}u^{\alpha}+\tau^{\alpha}\tau^{\gamma}u^{\beta})
\nonumber \\ &-&
\frac{1}{2}\Gamma_{2}(T\tau^{\mu}\sigma_{\mu})(h^{\alpha\beta}u^{\gamma}+h^{\beta\gamma}u^{\alpha}+h^{\gamma\alpha}u^{\beta})+\frac{1}{4}\Gamma'_{2}T\sigma{\mu}\tau_{\mu}u^{\alpha}u^{\beta}u^{\gamma}
\nonumber \\&-&
\frac{1}{2}\Gamma_{3}(\sigma^{2\langle\alpha\beta\rangle}u^{\gamma}+\sigma^{2\langle\beta\gamma\rangle}u^{\alpha}+\sigma^{2\langle\alpha\gamma\rangle}u^{\beta})+\frac{1}{2}(\Gamma_{2}+\Gamma'_{3})\sigma^{2\langle\mu\mu\rangle}u^{\alpha}u^{\beta}u^{\gamma}
\nonumber \\ &+&
\frac{1}{2}\Gamma_{3}\sigma^{2\langle\mu\mu\rangle}(h^{\alpha\beta}u^{\gamma}+h^{\beta\gamma}u^{\alpha}+h^{\gamma\alpha}u^{\beta})+\frac{1}{2}\Gamma'_{3}\sigma^{2\langle\mu\mu\rangle}u^{\alpha}u^{\beta}u^{\gamma}
\nonumber \\&+&
\frac{1}{2}\Gamma_{2}T(\sigma^{\langle\alpha\beta\rangle}\tau^{\gamma}+\sigma^{\langle\beta\gamma\rangle}\tau^{\alpha}+\sigma^{\langle\alpha\gamma\rangle}\tau^{\beta})
\nonumber \\&-&
\frac{1}{2}\Gamma'_{3}(\sigma^{\langle\gamma\mu\rangle}u^{\alpha}u^{\beta}\sigma_{\mu}+\sigma^{\langle\alpha\mu\rangle}u^{\beta}u^{\gamma}\sigma_{\mu}+\sigma^{\langle\beta\mu\rangle}u^{\alpha}u^{\gamma}\sigma_{\mu})
\nonumber \\ &+&
\frac{1}{2}\Gamma'_{2}T(u^{\alpha}u^{\beta}\sigma^{\langle\gamma\mu\rangle}\tau_{\mu} + u^{\beta}u^{\gamma}\sigma^{\langle\alpha\mu\rangle}\tau_{\mu} + u^{\gamma}u^{\alpha}\sigma^{\langle\beta\mu\rangle}\tau_{\mu})
\nonumber \\&+& \bigg(\Gamma_{3}-\frac{1}{2}\Gamma'_{3} \bigg)(u^{\alpha}u^{\beta}\sigma^{\langle\gamma\mu\rangle}\sigma_{\mu} + u^{\beta}u^{\gamma}\sigma^{\langle\alpha\mu\rangle}\sigma_{\mu} + u^{\gamma}u^{\alpha}\sigma^{\langle\beta\mu\rangle}\sigma_{\mu})
\nonumber \\ &-&
\frac{1}{2}\Gamma_{2}T(g^{\alpha\beta}\sigma^{\langle\gamma\mu\rangle}\tau_{\mu}+g^{\beta\gamma}\sigma^{\langle\alpha\mu\rangle}\tau_{\mu}+g^{\gamma\alpha}\sigma^{\langle\beta\mu\rangle}\tau_{\mu})
\nonumber \\&+&
\frac{1}{2}\Gamma_{3}(g^{\alpha\beta}\sigma^{\langle\gamma\mu\rangle}\sigma_{\mu}+g^{\beta\gamma}\sigma^{\langle\alpha\mu\rangle}\sigma_{\mu}+g^{\gamma\alpha}\sigma^{\langle\beta\mu\rangle}\sigma_{\mu})
\nonumber \\ &+&
\frac{1}{2}\Gamma_{2}(\sigma^{\langle\alpha\beta\rangle}u^{\gamma}+\sigma^{\langle\beta\gamma\rangle}u^{\alpha}+\sigma^{\langle\alpha\gamma\rangle}u^{\beta})(-T\tau)+\frac{3}{4}\Gamma'_{3}(\sigma^{\langle\alpha\beta\rangle}u^{\gamma}+\sigma^{\langle\beta\gamma\rangle}u^{\alpha}+\sigma^{\langle\alpha\gamma\rangle}u^{\beta})(\sigma)
\nonumber\\&+&
\frac{1}{2}\Gamma'_{2}T^{2}\tau(u^{\alpha}u^{\beta}\tau^{\gamma} + u^{\beta}u^{\gamma}\tau^{\alpha} + u^{\gamma}u^{\alpha}\tau^{\beta})-\frac{1}{2}\Gamma'_{3} \sigma(u^{\alpha}u^{\beta}\sigma^{\gamma} + u^{\beta}u^{\gamma}\sigma^{\alpha} + u^{\gamma}u^{\alpha}\sigma^{\beta})
\nonumber \\ &-&
\frac{1}{2}\Gamma_{2}T^{2}\tau(g^{\alpha\beta}\tau^{\gamma}+g^{\beta\gamma}\tau^{\alpha}+g^{\gamma\alpha}\tau^{\beta})
+
\frac{1}{2}\Gamma_{3}\sigma(g^{\alpha\beta}\sigma^{\gamma}+g^{\beta\gamma}\sigma^{\alpha}+g^{\gamma\alpha}\sigma^{\beta})
\bigg\}.
\eea
This allow us to express triple tensor up to quadratic in terms of $\pi, q^{\alpha}, t^{\langle\alpha\beta\rangle}$ which we provided in subsection.~\ref{subsec:constitutive}.
\section{Transport Coefficients}\label{app:transportc}
In this section, we provide the transport coefficients, viz. bulk viscosity $\zeta$, thermal conductivity $\kappa$ and shear viscosity $\eta$. Which reads
\bea\label{eq:transc}
\zeta &=& \frac{1}{2T}\frac{1}{\mbp}\frac{\begin{bmatrix}
\ddot{p} & \dot{p}-\dot{p}' & \dot{\Gamma}_{1} \\
\dot{p}-\dot{p}' & {p}'-{p}'' & \Gamma^{'}_{1}-\Gamma_{1}
\\
-\dot{p} & -{p}' & \frac{5}{3}{\Gamma}_{1}
\end{bmatrix}}{ 
\begin{bmatrix}
\ddot{p} & \dot{p}-\dot{p}'\\
\dot{p}-\dot{p}' & {p}'-{p}''
\end{bmatrix}}= -\frac{1}{\mbp}\frac{5}{6}\frac{nk_{B}^{3}T^{3}}{m^{2}}
\nonumber\\
\kappa &=& \frac{1}{2T}\frac{1}{\mbq}\begin{bmatrix}
\dot{p} & -\dot{\Gamma}_{1}\\
{p}' & \Gamma_{1}-\dot{\Gamma}_{1}
\end{bmatrix}\frac{1}{\dot{p}}= -\frac{1}{\mbq}\frac{5}{2}\frac{nk_{B}^{2}T^{2}}{m}
\\
\eta &=& \frac{1}{T}\frac{1}{\mbt}\Gamma_{1}= -\frac{1}{\mbt}nk_{B}T,\nonumber
\eea
where the second expressions on the right hand sides result from the equation of state was introduced in \cite{AIHPA_1997__67_2_111_0,RET} we also provide the expression below, in the limit of the non-relativistic gas, i.e. for $mc^{2} \gg k_{B}T$.The state function $p(\phi, T)$ for relativistic gases is given by the equation
\bea\label{eq:EOSp}
p &=& \text{exp}(-\frac{k_{B}}{m}\phi) 4\pi ym^{2}(k_{B}T)^{2}K_{2}\bigg(\frac{m}{k_{B}T}\bigg)\nonumber\\
&\implies& y m^{3}k_{B}T\bigg(2\pi\frac{k_{B}}{m}T\bigg)\text{exp}\bigg(-\frac{m}{k_{B}}\phi-\frac{m}{k_{B}T}\bigg)\bigg[1+\frac{15}{8}\frac{k_{B}T}{m}+ \cdots \bigg],
\eea
$k_{B}$ is the Boltzmann constant and $y$ is related to Planck constant. 
\subsection{Coefficients in non-degenerate regime}\label{app:coenondegen}
Here we define the values for coefficients $\mathcal{A}^{2}_{1}$ through to $\mathcal{A}^{2}_{9}$ and also $\mathcal{S}^{3}_{1}$ through to $\mathcal{S}^{3}_{10}$ for a non-degenerate limit and we obtain
\begin{eqnarray}\label{app:nondegfluxtcoef2}
\mathcal{A}^{2}_{1} &=&-\frac{18}{c^{2}}\frac{ \bigg(1-\frac{20}{\gamma^{2}}\bigg)+\frac{35}{\gamma^{3}}G -\bigg(1+\frac{49}{\gamma^{2}} \bigg)G^{2}-\frac{7}{\gamma}G^{3} +2G^{4}}{\frac{13}{\gamma}-\bigg(1-\frac{25}{\gamma}\bigg)G-\frac{17}{\gamma}G^{2}+3G^{3} +G^{4}},\nonumber\\
\mathcal{A}^{2}_{2} &=&-\frac{2}{\gamma}\frac{1+\frac{6}{\gamma}G-3G^{2} }{1+\frac{5}{\gamma}G-5G^{2} },\quad
\mathcal{A}^{2}_{3} =-\frac{1}{\gamma}\frac{-1-\frac{6}{\gamma}G+7G^{2}-G^{3}}{1+\frac{5}{\gamma}G-5G^{2}-G^{3}},\nonumber\\
\mathcal{A}^{2}_{4} &=&-\frac{3}{\gamma^{2}}\frac{1+\frac{6}{\gamma}G-G^{2} }{1+\frac{5}{\gamma}G-5G^{2} },\\
\mathcal{A}^{2}_{5}&=& \frac{3}{c^{2}}\bigg[\frac{\bigg(2-\frac{5}{\gamma^{2}}\bigg)+\frac{35}{\gamma}G-\bigg(2-\frac{22}{\gamma^{2}}\bigg)G^{2}}{ \frac{3}{\gamma} -\bigg(2-\frac{20}{\gamma^{2}}\bigg)G-\frac{13}{\gamma}G^{2}}+\frac{\bigg(1-\frac{6}{\gamma}\bigg)+\bigg(\frac{18}{\gamma}-\frac{27}{\gamma^{2}}\bigg)G-\frac{15}{\gamma^{2}}G^{2}}{ \frac{3}{\gamma} -\frac{36}{\gamma^{2}}G-\frac{15}{\gamma}G^{2}}\bigg],\nonumber\\
\mathcal{A}^{2}_{6}&=&\bigg(\frac{18}{\gamma}+\frac{3}{G}\bigg)^{2}, \quad \mathcal{A}^{2}_{7}=\frac{1}{G}\bigg(\frac{6}{\gamma^{2}}+\frac{3}{G}\bigg)^{2},\nonumber
\end{eqnarray}
and
\bea\label{app:nondegentcoe}
\Hc^{3}_{1} &=&\frac{k_{B}}{4m^{3}c^{2}}\frac{\gamma^{2}}{n^{2}}\frac{1-\frac{2}{\gamma^{2}}+ \frac{6}{\gamma}G-2G^{2}+G^{3}}{-\frac{13}{\gamma}+\bigg(1-\frac{25}{\gamma}\bigg)G-\frac{17}{\gamma}G^{2}+3G^{3}+G^{4}},\nonumber \\
\Hc^{3}_{2} &=& \frac{k_{B}}{2m^{2}c^{6}}\frac{\gamma}{n^{2}}\frac{\frac{6}{\gamma}\bigg(\frac{6}{\gamma^{2}}-3\bigg)G-\frac{16}{\gamma}G^{2}+G^{3} }{\bigg(1+\frac{5}{\gamma}G-5G^{2}\bigg)^{2}\bigg(1+\frac{5}{\gamma}G-5G^{2}-G^{3}\bigg)}, \nonumber \\
\Hc^{3}_{3} &=& \frac{k_{B}}{4m^{2}c^{4}}\frac{\gamma}{n^{2}} \frac{3 -\frac{5}{\gamma^{2}}+\bigg(\frac{18}{\gamma}-\frac{20}{\gamma^{3}}\bigg)G-\bigg(8-\frac{35}{\gamma^{2}}\bigg)G^{2}-\frac{16}{\gamma}G^{3}-2G^{4}+G^{5} }{\bigg(1+\frac{5}{\gamma}G-5G^{2}\bigg)\bigg[ -\frac{13}{\gamma}+\bigg(1-\frac{25}{\gamma}\bigg)G-\frac{17}{\gamma}G^{2}+3G^{3}+G^{4}\bigg] }, \nonumber\\
\Hc^{3}_{4} &=& -\frac{k_{B}}{4m^{2}c^{4}}\frac{\gamma^{2}}{n^{2}}\frac{1}{G}\bigg(1+\frac{5}{\gamma}G-G^{2}\bigg),\quad \Hc^{3}_{5} = -\frac{k_{B}}{m^{2}c^{4}}\frac{\gamma^{2}}{n^{2}}\frac{1}{G^{2}}\frac{1+\frac{6}{\gamma}G-\frac{1}{\gamma^{1}}-G^{2}}{1+\frac{5}{\gamma}G-G^{2}}, \\
\Hc^{3}_{6} &=&\frac{k_{B}}{4m^{3}c^{2}}\frac{\gamma^{2}}{n^{2}}\frac{-2+\frac{6}{\gamma^{2}}-\frac{30}{\gamma}G-\frac{22}{\gamma^{2}}G^{2}+G^{2} }{-\frac{13}{\gamma}+\bigg(1-\frac{25}{\gamma}\bigg)G-\frac{17}{\gamma}G^{2}+3G^{3}+G^{4}},\nonumber \\
\Hc^{3}_{7} &=&\frac{k_{B}}{2m^{3}c^{6}}\frac{\gamma}{n^{2}}\frac{\frac{6}{\gamma}-\bigg(1-\frac{12}{\gamma}\bigg)G+\frac{5}{\gamma}G^{2}+G^{3} }{\bigg(1+\frac{5}{\gamma}G-5G^{2}\bigg)\bigg(1+\frac{5}{\gamma}G-5G^{2}-G^{3}\bigg)},\quad
\Hc^{3}_{8} =\frac{k_{B}}{4m^{3}c^{2}}\frac{\gamma^{2}}{n^{2}}\frac{1}{G^{2}}\bigg(1+\frac{6}{\gamma}G\bigg)^{2},\nonumber
\eea
\begin{figure}[!ht]
	\centering
	\includegraphics[scale=0.75]{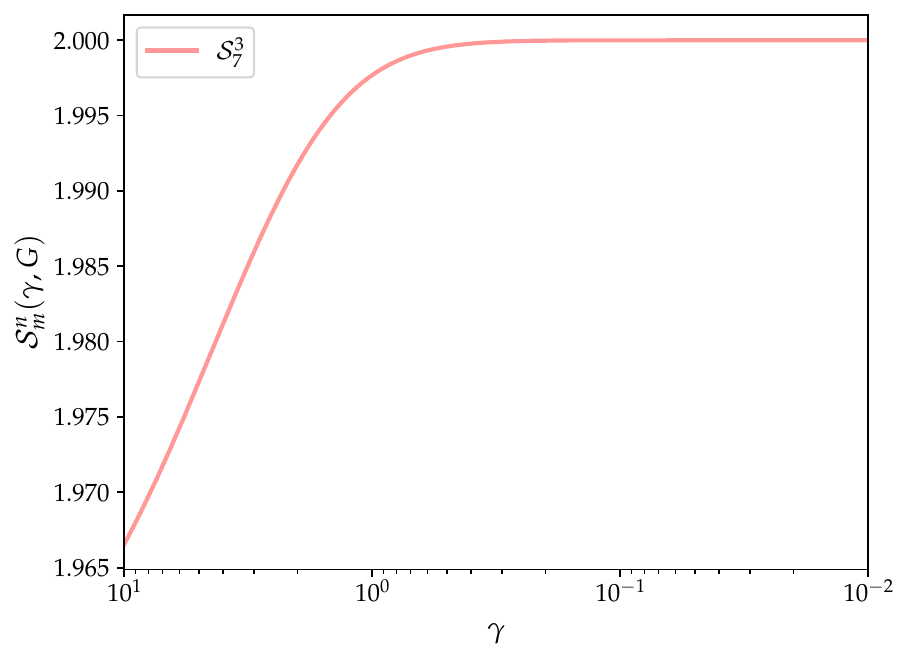}
	\caption{Plots of $\Hc^{i}_{j}$ coefficient as a function of the temperature.}
	\label{fig:s37coe}
\end{figure}	
\bea\label{app:nondegentcoe1}
\Hc^{3}_{9} &=& \frac{k_{B}}{4m^{2}c^{4}}\frac{\gamma}{n^{2}}\frac{1}{G} \frac{3 -\frac{5}{\gamma^{2}}+\bigg(\frac{18}{\gamma}-\frac{20}{\gamma^{3}}\bigg)G-\bigg(8-\frac{35}{\gamma^{2}}\bigg)G^{2}-\frac{16}{\gamma}G^{3}-2G^{4}+G^{5} }{\bigg(1+\frac{5}{\gamma}G-5G^{2}\bigg)\bigg[ -\frac{13}{\gamma}+\bigg(1-\frac{25}{\gamma}\bigg)G-\frac{17}{\gamma}G^{2}+3G^{3}+G^{4}\bigg] }, \nonumber\\
\Hc^{3}_{10} &=& \frac{18k_{B}}{16m^{2}c^{4}}\frac{\gamma}{n^{2}}\frac{1+\frac{6}{\gamma}G-\frac{5}{\gamma }G-G^{2}}{G\bigg(1+\frac{5}{\gamma}G-G^{2} \bigg)}.\nonumber
\eea
 

\bibliographystyle{JHEP}
\bibliography{RelativisticThermo}

@ARTICLE{Juttner1911-ou, 
title = "Das Maxwellsche Gesetz der Geschwindigkeitsverteilung in der Relativtheorie", author = "J{\"u}ttner, Ferencz", journal = "Ann. Phys.", publisher = "Wiley", volume = 339, number = 5, pages = "856--882", year = 1911, language = "de" 
}

@article{Eckart,
  title = {The Thermodynamics of Irreversible Processes. III. Relativistic Theory of the Simple Fluid},
  author = {Eckart, Carl},
  journal = {Phys. Rev.},
  volume = {58},
  issue = {10},
  pages = {919--924},
  numpages = {0},
  year = {1940},
  month = {Nov},
  publisher = {American Physical Society},
  doi = {10.1103/PhysRev.58.919},
  url = {https://link.aps.org/doi/10.1103/PhysRev.58.919}
}

@article{Chung2017,
author = {Sae Rom Chung and Chong Hyun Suh and  Choi and Jeong Hyun Lee},
title = {Safety of radiofrequency ablation of benign thyroid nodules and recurrent thyroid cancers: a systematic review and meta-analysis},
journal = {International Journal of Hyperthermia},
volume = {33},
number = {8},
pages = {920-930},
year  = {2017},
publisher = {Taylor & Francis},
doi = {10.1080/02656736.2017.1337936},
    note ={PMID: 28565997},

URL = { 
        https://doi.org/10.1080/02656736.2017.1337936
    
},
eprint = { 
        https://doi.org/10.1080/02656736.2017.1337936
    
}

}

@BOOK{Vincent1965,
       author = {{Vincenti}, Walter Guido and {Kruger}, Charles H.},
        title = "{Introduction to physical gas dynamics}",
        publisher = "{}", 
         year = 1965,
       adsurl = {https://ui.adsabs.harvard.edu/abs/1965itpg.book.....V}
}

@BOOK{Zeldovich1967,
       author = {{Zel'dovich}, Ya. B. and {Raizer}, Yu. P.},
        title = "{Physics of shock waves and high-temperature hydrodynamic phenomena}",
        publisher = "Springer", 
         year = 1967,
       adsurl = {https://ui.adsabs.harvard.edu/abs/1967pswh.book.....Z}
}

@book{Groot1984,
	address = {New York},
	edition = {Dover ed},
	title = {Non-equilibrium thermodynamics},
	isbn = {9780486647418},
	publisher = {Dover Publications},
	author = {Groot, S. R. de and Mazur, P.},
	year = {1984},
}

@MASTERSTHESIS{fhumulani2010,
  AUTHOR =       {Fhumulani Nemulodi},
  TITLE =        {Third order relativistic dissipative 
fluid dynamics for
heavy-ion collisions},
  SCHOOL =       {University of Cape Town},
  YEAR =         {2010},
  month =        {December},
  }

@book{DeGroot:1980dk,
    author = "De Groot, S. R.",  
    title = "{Relativistic Kinetic Theory. Principles and Applications}",
    year = "1980",
publisher = "Wiley"
}

@misc{Ralph2104,
  doi = {10.48550/ARXIV.2104.06965},
  
  url = {https://arxiv.org/abs/2104.06965},
  
  author = {Chamberlin, Ralph V. and Clark, Michael R. and Mujica, Vladimiro and Wolf, George H.},
  
  title = {Multiscale Thermodynamics: Energy, Entropy, and Symmetry from Atoms to Bulk Behavior},
  
  publisher = {arXiv},
  
  year = {2021},
  
  copyright = {Creative Commons Attribution 4.0 International}
}

@article{Matyas2021,
	doi = {10.1515/jnet-2021-0022},
  
	url = {https://doi.org/10.1515%2Fjnet-2021-0022},
  
	year = 2021,
	month = {sep},
  
	publisher = {Walter de Gruyter {GmbH}
},
  
	volume = {47},
  
	number = {1},
  
	pages = {31--60},
  
	author = {M{\'{a}}ty{\'{a}}s Szücs and Michal Pavelka and R{\'{o}}bert Kov{\'{a}}cs and Tam{\'{a}}s Fülöp and P{\'{e}}ter V{\'{a}}n and Miroslav Grmela},
  
	title = {A Case Study of Non-Fourier Heat Conduction Using Internal Variables and {GENERIC}},
  
	journal = {Journal of Non-Equilibrium Thermodynamics}
}

@Article{Cimmelli:2014,
AUTHOR = {Cimmelli, Vito Antonio and Jou, David and Ruggeri, Tommaso and Ván, Péter},
TITLE = {Entropy Principle and Recent Results in Non-Equilibrium Theories},
JOURNAL = {Entropy},
VOLUME = {16},
YEAR = {2014},
NUMBER = {3},
PAGES = {1756--1807},
URL = {https://www.mdpi.com/1099-4300/16/3/1756},
ISSN = {1099-4300},
DOI = {10.3390/e16031756}
}

@inproceedings{Asproulis:2008,
title = "A hybrid molecular continuum method using point wise coupling",
author = "N. Asproulis and M. Kalweit and D. Drikakis",
note = "Copyright: Copyright 2012 Elsevier B.V., All rights reserved.; 6th International Conference on Engineering Computational Technology, ECT 2008 ; Conference date: 02-09-2008 Through 05-09-2008",
year = "2008",
language = "English",
isbn = "9781905088249",
series = "Proceedings of the 6th International Conference on Engineering Computational Technology",
booktitle = "Proceedings of the 6th International Conference on Engineering Computational Technology",
}

@article{Israel:1979wp,
    author = "Israel, W. and Stewart, J. M.",
    title = "{Transient relativistic thermodynamics and kinetic theory}",
    doi = "10.1016/0003-4916(79)90130-1",
    journal = "Annals Phys.",
    volume = "118",
    pages = "341--372",
    year = "1979"
}

@ARTICLE{Israel_1976,
       author = {{Israel}, W. and {Stewart}, J.~M.},
        title = "{Thermodynamics of nonstationary and transient effects in a relativistic gas}",
      journal = {Physics Letters A},
         year = 1976,
        month = sep,
       volume = {58},
       number = {4},
        pages = {213-215},
          doi = {10.1016/0375-9601(76)90075-X},
       adsurl = {https://ui.adsabs.harvard.edu/abs/1976PhLA...58..213I}
}

@article{Younus:2019rrt,
    author = "Younus, Mohammed and Muronga, Azwinndini",
    title = "{Third order viscous hydrodynamics from the entropy four current}",
    eprint = "1910.11735",
    archivePrefix = "arXiv",
    primaryClass = "nucl-th",
    doi = "10.1103/PhysRevC.102.034902",
    journal = "Phys. Rev. C",
    volume = "102",
    number = "3",
    pages = "034902",
    year = "2020"
}

@article{Muronga_2008,
  title = {Relativistic dynamics of nonideal fluids: Viscous and heat-conducting fluids. I. General aspects and $3+1$ formulation for nuclear collisions},
  author = {Muronga, Azwinndini},
  journal = {Phys. Rev. C},
  volume = {76},
  issue = {1},
  pages = {014909},
  numpages = {20},
  year = {2007},
  month = {Jul},
  publisher = {American Physical Society},
  doi = {10.1103/PhysRevC.76.014909},
  url = {https://link.aps.org/doi/10.1103/PhysRevC.76.014909}
}

@article{Muronga_2010,
	doi = {10.1088/0954-3899/37/9/094008},
	url = {https://doi.org/10.1088/0954-3899/37/9/094008},
	year = 2010,
	month = {aug},
	publisher = {{IOP} Publishing},
	volume = {37},
	number = {9},
	pages = {094008},
	author = {Azwinndini Muronga},
	title = {New developments in relativistic dissipative fluid dynamics},
	journal = {Journal of Physics G: Nuclear and Particle Physics}
}

@article{Ruggeri2005,
author = {Ruggeri, Tommaso},
journal = {Bollettino dell'Unione Matematica Italiana},
language = {eng},
month = {2},
number = {1},
pages = {1-20},
publisher = {Unione Matematica Italiana},
title = {The entropy principle: from continuum mechanics to hyperbolic systems of balance laws},
url = {http://eudml.org/doc/196153},
volume = {8-B},
year = {2005},
}

@book{HIST,
author = {Müller, Ingo},
year = {2007},
month = {01},
pages = {},
publisher ={Springer}, 
title = {A History of Thermodynamics. The Doctrine of Energy and Entropy},
journal = {A History of Thermodynamics: The Doctrine of Energy and Entropy},
doi = {10.1007/978-3-540-46227-9}
}

@book{RET,
author = {Müller, Ingo and Ruggeri, Tommaso},
year = {1998},
month = {01},
pages = {},
title = {Rational Extended Rational Thermodynamics},
publisher = {Springer},
volume = {37},
isbn = {978-1-4612-7460-5},
doi = {10.1007/978-1-4612-2210-1}
}

@book{Misner:1974qy,
    author = "Misner, Charles W. and Thorne, K. S. and Wheeler, J. A.",
    title = "{Gravitation}",
    isbn = "978-0-7167-0344-0, 978-0-691-17779-3",
    publisher = "W. H. Freeman",
    address = "San Francisco",
    year = "1973"
}

@article {Liu1972,
    AUTHOR = {Liu, I Shih},
     TITLE = {Method of {L}agrange multipliers for exploitation of the
              entropy principle},
   JOURNAL = {Arch. Rational Mech. Anal.},
  FJOURNAL = {Archive for Rational Mechanics and Analysis},
    VOLUME = {46},
      YEAR = {1972},
     PAGES = {131--148},
      ISSN = {0003-9527},
   MRCLASS = {80.35},
  MRNUMBER = {337164},
MRREVIEWER = {Thomas O. Philips},
       DOI = {10.1007/BF00250688},
       URL = {https://doi.org/10.1007/BF00250688},
}

@ARTICLE{Liu:1986anpl, 
       author = {{Liu}, I. -Shih and {M{\"u}ller}, I. and {Ruggeri}, T.},
        title = "{Relativistic thermodynamics of gases}",
      journal = {Annals of Physics},
         year = 1986,
        month = jun,
       volume = {169},
       number = {1},
        pages = {191-219},
          doi = {10.1016/0003-4916(86)90164-8},
       adsurl = {https://ui.adsabs.harvard.edu/abs/1986AnPhy.169..191L}
}

@article{Friedrichs1971aa,
    author = "K. O. Friedrichs and P. D. Lax." , 
    title = "Systems of conservation equations with a convex extension.", 
    doi = "10.1073/pnas.68.8.1686",
    journal ="Proc. Bat. Acad. Sci. USA", 
    volume="68",
    pages ="1686-1688",
    number ="8",
    year ="1971"
}

@article{Geroch:1990bw,
    author = "Geroch, Robert P. and Lindblom, L.",
    title = "{Dissipative relativistic fluid theories of divergence type}",
    doi = "10.1103/PhysRevD.41.1855",
    journal = "Phys. Rev. D",
    volume = "41",
    pages = "1855",
    year = "1990"
}

@article{Calzetta:1997aj,
    author = "Calzetta, Esteban",
    title = "{Relativistic fluctuating hydrodynamics}",
    eprint = "gr-qc/9708048",
    archivePrefix = "arXiv",
    doi = "10.1088/0264-9381/15/3/015",
    journal = "Class. Quant. Grav.",
    volume = "15",
    pages = "653--667",
    year = "1998"
}

@article{AIHPA_1997__67_2_111_0,
     author = {Kremer, G. M. and M\"uller, Ingo},
     title = {Dynamic pressure in relativistic thermodynamics},
     journal = {Annales de l'I.H.P. Physique th\'eorique},
     pages = {111--121},
     publisher = {Gauthier-Villars},
     volume = {67},
     number = {2},
     year = {1997},
     zbl = {0885.76088},
     mrnumber = {1472564},
     language = {en},
     url = {http://www.numdam.org/item/AIHPA_1997__67_2_111_0/}
}

@article{AIHPA_1998__69_3_309_0,
     author = {Kremer, G. M. and M\"uller, Ingo},
     title = {Thermal conductivity and dynamic pressure in extended thermodynamics of chemically reacting mixture of gases},
     journal = {Annales de l'I.H.P. Physique th\'eorique},
     pages = {309--334},
     publisher = {Gauthier-Villars},
     volume = {69},
     number = {3},
     year = {1998},
     zbl = {0964.80007},
     mrnumber = {1648986},
     language = {en},
     url = {http://www.numdam.org/item/AIHPA_1998__69_3_309_0/}
}

@article{pennisi1986,
    author ="S. Pennisi",
    title  =" Some representation theorems in a 4-dimensional production space.",
    journal = "Fisica Matematica Suppl.",
    publisher="B.U.M.I.",
    volume ="5",
    year = "(1986)"
}

@article{smithgf1965,
     author = "G. F. Smith",
     title = " On Isotropic Integrity  Bases.",
     journal = " Arch. Rational. Mech. Anal.",
     volume ="18",
     year = "(1965)"
}

@article{wangcc1969,
     author ="Wang, C.C.",
     title = "On Representations for Isotropic Functions, Part I. Archive for Rational Mechanics and Analysis",  volume = "33" ,
     number = "249",
     doi= "10.1007/BF00281278",
     year = "(1969)"
}

@article{Salazar_2020,
    author = {J. FELIX SALAZAR and THOMAS ZANNIAS},
    title = {ON EXTENDED THERMODYNAMICS:FROM CLASSICAL TO THE RELATIVISTIC REGIME},
    eprint = "gr-qc/1904.04368",
    archivePrefix = "arXiv",
    journal = "Int.Journal.Modern.Phydics.D",
	doi = {10.1142/s0218271820300104},
  
	url = "{https://doi.org/10.1142\%2Fs0218271820300104}",
    month = "10",
	year = "2020",
	
}

@book {RS,
    AUTHOR = {Ruggeri, Tommaso and Sugiyama, Masaru},
     TITLE = {Classical and relativistic rational extended thermodynamics of
              gases},
 PUBLISHER = {Springer, Cham},
      YEAR = {[2021] \copyright 2021},
     PAGES = {xxxii+669},
      ISBN = {978-3-030-59143-4; 978-3-030-59144-1},
   MRCLASS = {80-03 (80A05 80A10 82C40 82D05)},
  MRNUMBER = {4260900},
       DOI = {10.1007/978-3-030-59144-1},
       URL = {https://doi.org/10.1007/978-3-030-59144-1},
}

@article {Boillat1997,
    AUTHOR = {Boillat, Guy and Ruggeri, Tommaso},
     TITLE = {Hyperbolic principal subsystems: entropy convexity and
              subcharacteristic conditions},
   JOURNAL = {Arch. Rational Mech. Anal.},
  FJOURNAL = {Archive for Rational Mechanics and Analysis},
    VOLUME = {137},
      YEAR = {1997},
    NUMBER = {4},
     PAGES = {305--320},
      ISSN = {0003-9527},
   MRCLASS = {82C40 (35L65 76P05)},
  MRNUMBER = {1463797},
MRREVIEWER = {Carlo Cercignani},
       DOI = {10.1007/s002050050030},
       URL = {https://doi.org/10.1007/s002050050030},
}

@book {LL,
    AUTHOR = {Landau, L. D. and Lifshitz, E. M.},
     TITLE = {Fluid mechanics},
    SERIES = {Course of Theoretical Physics, Vol. 6},
      NOTE = {Translated from the Russian by J. B. Sykes and W. H. Reid},
 PUBLISHER = {Pergamon Press, London-Paris-Frankfurt; Addison-Wesley
              Publishing Company, Inc., Reading, Mass.},
      YEAR = {1959},
     PAGES = {xii+536},
   MRCLASS = {76.00},
  MRNUMBER = {0108121},
}

@misc{https://doi.org/10.48550/arxiv.2102.02634,
  doi = {10.48550/ARXIV.2102.02634},
  
  url = {https://arxiv.org/abs/2102.02634},
  
  author = {Moloi, Teboho},
  
  title = {Spherical Bessel functions},
  
  publisher = {arXiv},
  
  year = {2021},
  
  copyright = {Creative Commons Attribution Share Alike 4.0 International}
}

\end{document}